\documentclass[12pt]{iopart}

\usepackage{graphicx}
\begin{document}

\title{Evidence for temporary and local transition of sp$^2$ graphite-type to sp$^3$ diamond-type bonding induced by the tip of an atomic force microscope} 

\author{Thomas Hofmann$^1$, Xinguo Ren$^2$, Alfred J. Weymouth$^1$, Daniel Meuer$^1$, Alexander Liebig$^1$, Andrea Donarini$^3$, Franz J. Giessibl$^{1}$}
 \ead{franz.giessibl@ur.de}
\address{$^1$Institute of Experimental and Applied Physics, Department of Physics, University of Regensburg, 93040 Regensburg, Germany}
\address{$^2$Beijing National Laboratory for Condensed Matter Physics, Institute of Physics, Chinese Academy of Sciences, 3rd South Str. 8,  Beijing 100190, China}
\address{$^3$Institute of Theoretical Physics, Department of Physics, University of Regensburg, 93040 Regensburg, Germany}

\date{\today}

\begin{abstract}
Artificial diamond is created by exposing graphite to pressures on the order of 10\,GPa and temperatures of about 2000\,K. Here, we provide evidence that the pressure exerted by the tip of an atomic force microscope onto graphene over the carbon buffer layer of silicon carbide can lead to a temporary transition of graphite to diamond on the atomic scale. We perform atomic force microscopy with CO terminated tips and copper oxide (CuOx) tips to image graphene and to induce the structural transition. A local transition induced by the force of the tip is accompanied by local rehybridization from an sp$^2$-bonded to an sp$^3$-bonded local structure. Density functional theory (DFT) predicts that a repulsive threshold of $\approx13$\,nN, followed by a force reduction by $\approx4$\,nN is overcome when inducing the graphite-diamond transition. The experimental observation of the third harmonic with a magnitude of about 200\,fm fits well to overcoming a force barrier of $F_{barrier}\approx 5$\,nN, followed by a force reduction by $-F_{barrier}$ and an upswing by $F_{barrier}$ for decreasing distances.  Experimental evidence for this transition is provided by the emergence of third harmonics in the cantilever oscillation when the laterally flexible CO terminated tip exerts a large repulsive force. Probing the sample with rigid CuOx tips in the strong repulisive regime shows a strong difference in the yielding of the A versus B sites to the pressure of the tip. The large repulsive overall force of $\approx 10$\,nN is only compatible with the experimental data if one assumes that the repulsive force acting on the tip when inducing the transition is compensated by a heavily increased van-der-Waals attraction of the tip due to form fitting of tip and sample by local indentation. 
The experiment also shows that atomic force microscopy allows to perform high pressure physics on the atomic scale.
\end{abstract}

\pacs{Valid PACS appear here}
\maketitle

\newpage
\tableofcontents

\newpage

\section{\label{sec:Introduction}Diamond versus graphite and transitions}
 
\begin{figure}
\begin{center}
\includegraphics[clip=true, width=0.7\textwidth]{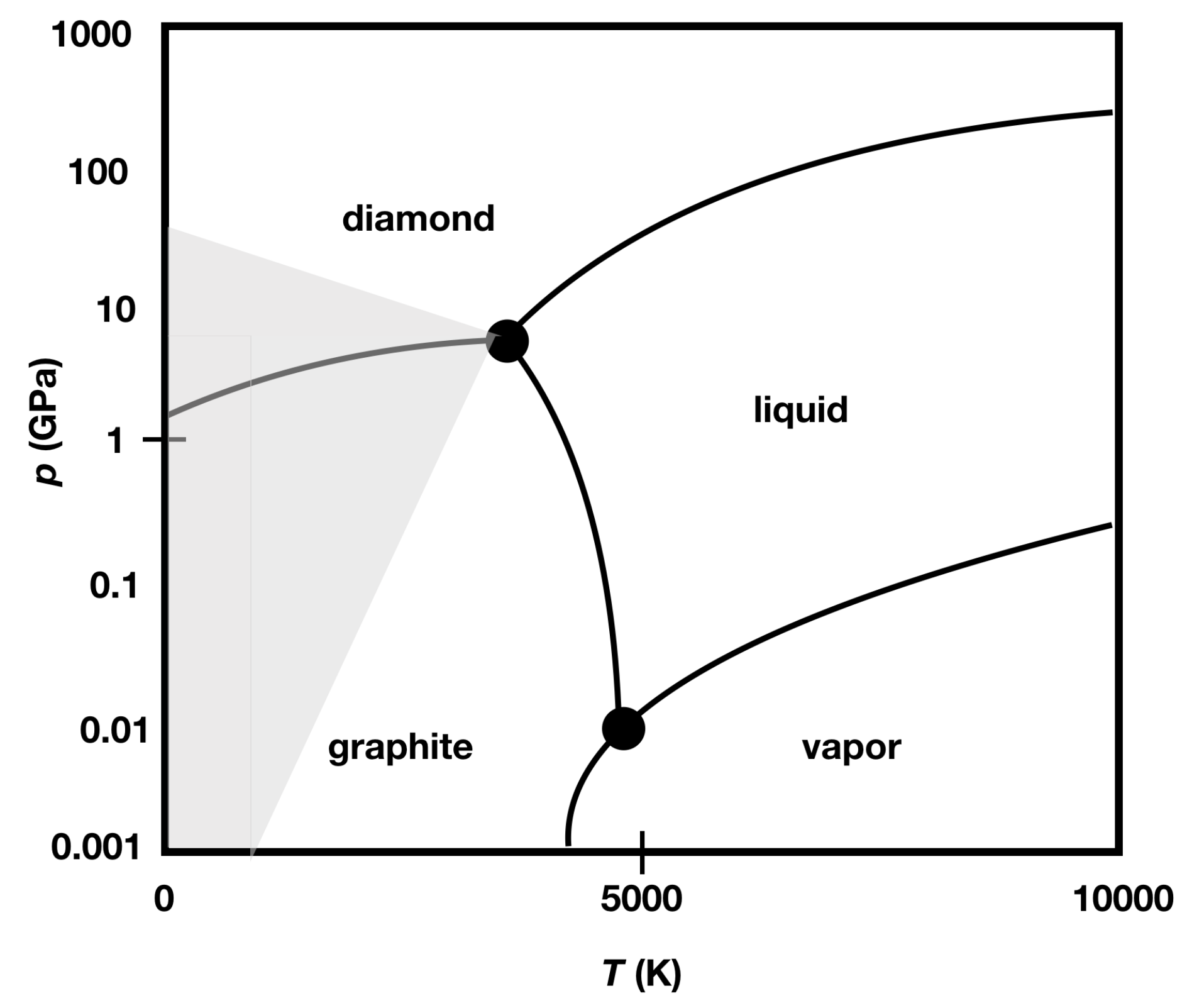}
\end{center}
\caption{Schematic phase diagram of carbon after data from Fig. 4 in \cite{Bundy1989}. The phase boundary between graphite and diamond for temperatures below 4000\,K is not a sharp line but a wide coexistence area indicated by the zone shaded in gray.}\label{fig01_carbon_phase_diagram}
\end{figure}

\lq\lq Deep within the earth diamonds grow. Diamonds the size of footballs, diamonds the size of watermelons -- countless billions of tons of diamonds wait for eternity a hundred miles beyond our reach.\rq\rq -- this is the opening sentence in chapter I of Robert M. Hazen's stunning account on the history of making artificial diamond \cite{Hazen1999} as reported in \cite{Bundy1955,Bovenkerk1959}. The pressure in the mantle of the earth at a depth of 100 miles is about 10\,GPa, and fractions of these football-sized diamonds can survive the trip to the earth's surface only when transported by volcanic eruptions of great speed \cite{Hazen1999}. Most of these \lq\lq water melon sized diamonds\rq\rq emerge on the surface of the earth as graphite, the stable phase of carbon for pressures below about 2\,GPa at low temperatures. A schematic phase diagram that shows the large coexistence area shaded in gray - both graphite and diamond are stable in ambient conditions, is presented in Fig. \ref{fig01_carbon_phase_diagram} after data from \cite{Bundy1989}.

Graphite and diamond coexist over a large range of pressures and temperatures, indicated by the gray zone in the phase diagram of Figure \ref{fig01_carbon_phase_diagram} created with data from \cite{Hazen1999,Bundy1989}. Therefore, transforming graphite to synthetic diamond requires to push the material through this wide gray area utilizing massive and ingenious reactor cells that sustain a pressure on the order of 10\,GPa at temperatures around 2000\,K \cite{Hazen1999}. At ambient conditions, graphite is slightly more stable than diamond by about 3\,kJ/mol or 30\,meV per atom \cite{Grochala2014,Popov2019}, while at 0\,K, diamond appears to be more stable than graphite by an even smaller energy  \cite{Grochala2014}. The two phases separated by a barrier of about 0.5\,eV per atom (Fig. 4a in \cite{Popov2019}), and it was already found in 1962 \cite{Howes1962} that diamond starts to transform into graphite both inside a crystal as well as on the surface by heating it in a vacuum to 1700\,$^{\circ}$C. This temperature corresponds to $k_B\cdot T \approx 172$\,meV. While this energy is significantly below the barrier of 0.5\,eV, the attempt frequency is expected to be around the frequency of optical phonons in diamond of about 40\,THz, yielding a significant transition rate even for thermal energies below the barrier.

Recently, the tip of an an atomic force microscope (AFM) \cite{Binnig1986} was used to induce a transition of bilayer graphene to diamond \cite{Gao2018} or 2D boron nitride to a diamond-like structure by AFM \cite{Cellini2021}. These nano indentation experiments applied loads of up to 100\,nN to a blunt spherical Si probe with, compared to atomic dimensions, a large contact area, lead to these sp$^2$ type bonding to diamond-like sp$^3$ type bonding conversions \cite{Gao2018}. Recently, such transistions have also been observed in boron nitrate on a SiO$_2$ substrate \cite{Cellini2021}.

Because of the increasing relative stability of matter with decreasing size, one might even expect that it is easier to induce this transition with an atomically sharp tip. The strength of materials with a given cross section is typically much smaller than the strength of an individual bond times the number of bonds that bridge the cross section. The reason is the slipping of bonds one by one under the influence of shear forces (see e.g. Vol. II, chapter 30-7 in \cite{Feynman2010}).

The tip of an AFM can create enormous pressure when its repulsive force is confined on a single unit cell. A surface unit cell in graphite spans an area of 246\,pm $\times$ 246\,pm $\times \sin(\pi/3)$ and forms a honeycomb mesh with two basis atoms spaced at 142\,pm. In AFM, forces between the front atom of the tip and the sample atom next to it can reach forces in the nanonewton regime without creating damage. A small repulsive force of merely 525\,pN over one unit cell already invokes a huge pressure of 10\,GPa. 

A transition of the bonding type between a CO terminated tip and a Fe adatom on Cu(111) has recently been observed \cite{Huber2019}. In this experiment, approaching the CO terminated tip to the Fe adatom resulted first in a local van-der-Waals minimum of about -10\,pN at a distance of 400\,pm, followed by a repulsive barrier of about +20\,pN at 300\,pm ending in a weak covalent bond with an absolute minimum of about -400\,pN at a distance of 260\,pm. Here, we attempt to modify the bonding character between the graphene surface layer and the buffer layer underneath on the scale of a few atoms with an atomically sharp tip.

\section{Density functional theory calculations on the effect of a high load of an atomic contact to bilayer graphene on SiC}
\label{sec:Density functional theory and proposed generation of higher harmonics}
 
The creation of diamond in pressure cells differs largely from exerting pressure by a single front atom of an AFM tip. Density functional theory (DFT) can simulate the effect of impinging an atomically sharp CO tip on bilayer graphene on SiC. For simplicity, we only consider the CO molecule pointing with the O atom towards the sample as the AFM tip.
DFT within the generalized gradient approximation of Perdew, Burke, and Ernzerhof \cite{Perdew1996}, as implemented in the Fritz-Haber-Institute-aims code package \cite{Blum2009} is used here to calculate the effect of a CO terminated tip that is located in the center of a graphene hexagon on SiC(0001). We build on a previous DFT study of this sample system by Nemec et al. \cite{Nemec2013}. The calculations were performed at zero bias voltage and a temperature of 0\,K.

\begin{figure}
\begin{center}
\includegraphics[clip=true, width=1\textwidth]{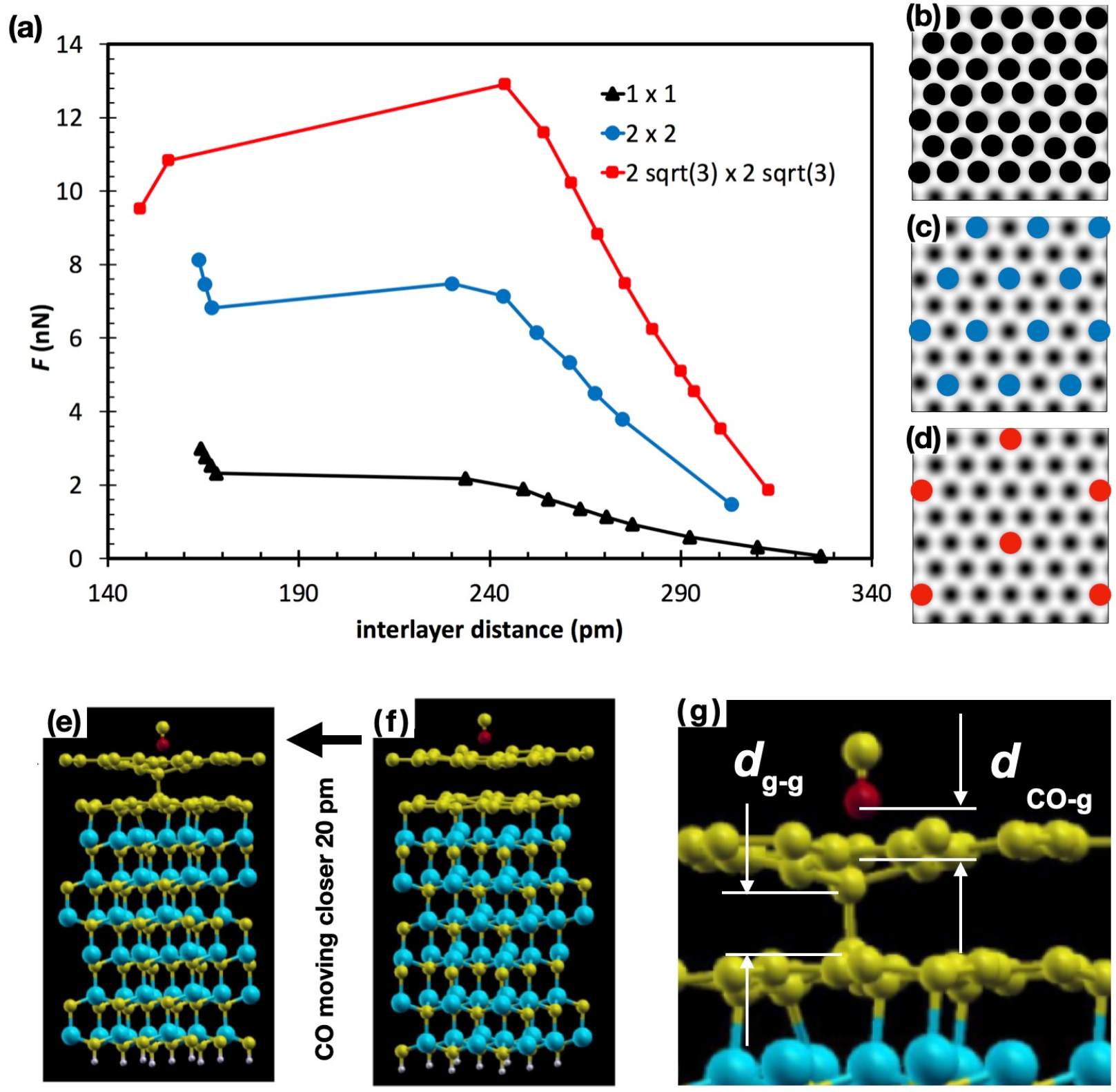}
\end{center}
\caption{A CO molecule is located over the hollow cite (center) of a hexagonal unit cell of bilayer graphene on SiC. (a) DFT calculated force spectra that occur during rehybridization for monolayer graphene on a graphene buffer layer on SiC as a function of interlayer distance $d_{g-g}$ as illustrated in (g). Three different sizes of unit cells are shown: (b) in the black curve, a CO molecule presses on every unit cell ($1\times 1$ structure), (c) the blue curve refers to one CO molecule for four unit cells ($2\times 2$ structure), and (d) finally one CO molecule for 12 unit cells ($2\sqrt{3}\times 2\sqrt{3}$ structure). (e) The slab is shown under high load where the transition has taken place. (f) Same slab with CO 20\,pm further away. (g) Close up of the tip region of (e): The CO tip molecule is centered over the carbon hexagon, and one of the three A sites of the graphene surface layer moves significantly towards the buffer layer and formes a covelent bond with a A$_{BL}$ atom.  }\label{fig02_DFT_g_d_trans_uc_dep}
\end{figure}
Figure \ref{fig02_DFT_g_d_trans_uc_dep} shows the results of a DFT calculation where a CO terminated tip is placed above the center of a graphene hexagon and the repulsive force on the CO tip termination is plotted against the interlayer distance between the monolayer graphene and the carbon buffer layer for three different sizes of unit cells. 

The black curve corresponds to a $1\times 1$ structure where a CO terminated tip is placed over each unit cell as indicated in Figure \ref{fig02_DFT_g_d_trans_uc_dep} (b). The force starts at zero at the equilibrium interlayer distance of 335\,pm and then rises sharply to a value of about 2\,nN where it remains quite flat over a nearest inter-graphene-layer distance from 235\,pm down to about 160\,pm, close to the bond length in diamond, where the force rises sharply again for even smaller distances.

The blue curve refers to a $2\times 2$ structure, corresponding to one CO tip in four unit cells as depicted in Figure \ref{fig02_DFT_g_d_trans_uc_dep} (c). The force again starts at zero at the equilibrium interlayer distance of 335\,pm and then rises even more sharply to a value of about 7\,nN where it declines until the nearest inter-graphene-layer distance reaches 160\,pm, again increasing profoundly upon further distance reduction.

Finally, the red curve refers to a $2\sqrt{3}\times 2\sqrt{3}$ structure, corresponding to one CO tip in twelve unit cells. Here, the force even rises up to a value of about 13\,nN before it declines again until a distance of 150\,pm. The calculated curve does not cover the even smaller distances that are expected to show another rising repulsive force.

It should be noted, however, to exert such a high force (pressure) from the CO tip, in our simulations, the CO molecule has to be kept vertically. Without doing so, the CO molecule will be tilted away from the upright position before reaching such a high force. How the CO tip might be stabilized under the experimental condition will be discussed further below.

\begin{center}
\begin{table}
\begin{tabular}{|c||c|c|c|c|c|}
    \hline
    number of                   &  surface & $F_{rehybridization}$  &$F_{rehybridization}$                 & pressure          \\
   unit cells                      &    vectors        & eV/\AA                &  nN                                 &     GPa            \\
    \hline \hline
    1                               &       $1\times 1$                         &   1.33  & 2                & 41                                 \\
    \hline
    4                               &       $2\times 2$                         &   4.7  &  7                & 36                                \\
    \hline
   12                              &       $2\sqrt{3}\times 2\sqrt{3}$  &   8.0     & 13               & 20                                 \\
    \hline
\end{tabular}
\caption{Rehybridization thresholds as a function of unit cell size.}\label{table:rehybridization_thresholds}
\end{table}
\end{center}

While the increase of unit cell sizes has led to increased force thresholds that induce the graphite to diamond transition, we expect that the force threshold saturates for large enough unit cells. Experimental and theoretical evidence shows that when graphene is exposed to forces from the tip of a scanning probe microscope close by, the largest part of the bending occurs within nanometers from the tip \cite{Klimov2012}.

Table \ref{table:rehybridization_thresholds} summarizes the results of Fig.\ref{fig02_DFT_g_d_trans_uc_dep} and lists the corresponding average pressures. As expected, the average pressure falls for larger unit cells. However, the calculated pressure where a transition occurs for the $1\times 1$ structure a t$T=0$\,K is at 41\,GPa surprisingly high, given that the empirically found pressure needed for artificial diamond creation is about 10\,GPa.

A possible reason for this discrepancy is provided by Grochala, who notes that thermal activation is important in particular when transforming graphite to diamond, \lq\lq which results in immeasurably sluggish transformation at $T<1000$\,K even when the pressures involved are larger than necessary for transformation\rq\rq (p.3682 in \cite{Grochala2014}).

Although the sample is at a temperature of about 4\,K in our experiment, it is conceivable that the oscillating motion of our tip with its amplitude of $A=50$\,pm shakes the lattice in a similar fashion as thermal excitation and thus helps the atoms to find their new equilibrium positions as the thermal motion at the high temperatures requried in artificial diamond creation. One may argue that the oscillation of the cantlever is very slow in frequency (40\,kHz) with respect to optical phonon frequencies (40\,THz) and should therefore considered to be adiabatic. However, as we will see below, dissipation is observed in the experiment, proving that the oscillation of the tip to and from the sample is not a conservative process, but runs through a hysteresis loop as explained by Sasaki and Tsukada (Fig. 2) \cite{Sasaki2000}. The jumps at the edges of the hysteresis loop are limited by phonon dynamics and are therefore very similar to direct thermal excitation.

The equipartion theorem predicts that the thermal motion of an atom with mass $m_a$ held by bonds with an atomic stiffness $k_{bond}$ has a thermally excited motion $\delta x$ in every spatial degree of freedom of
\begin{equation}\label{eq_thermal_motion}
 \frac{1}{2}k_{bond} <\delta x^2> = \frac{1}{2} k_{B} T,
\end{equation}
where $k_{B} =1.38\cdot 10^{-23}$\,J/K is Boltzmann's constant.
At a temperature of $T\approx 2000$\,K and an estimated value for $k_{bond}\approx 300$\,N/m, a total thermal motion of about 10\,pm is estimated. This motion will certainly help to find the new local energy minimum that the diamond structure provides under high pressure.

\begin{figure}
\begin{center}
\includegraphics[clip=true, width=0.6\textwidth]{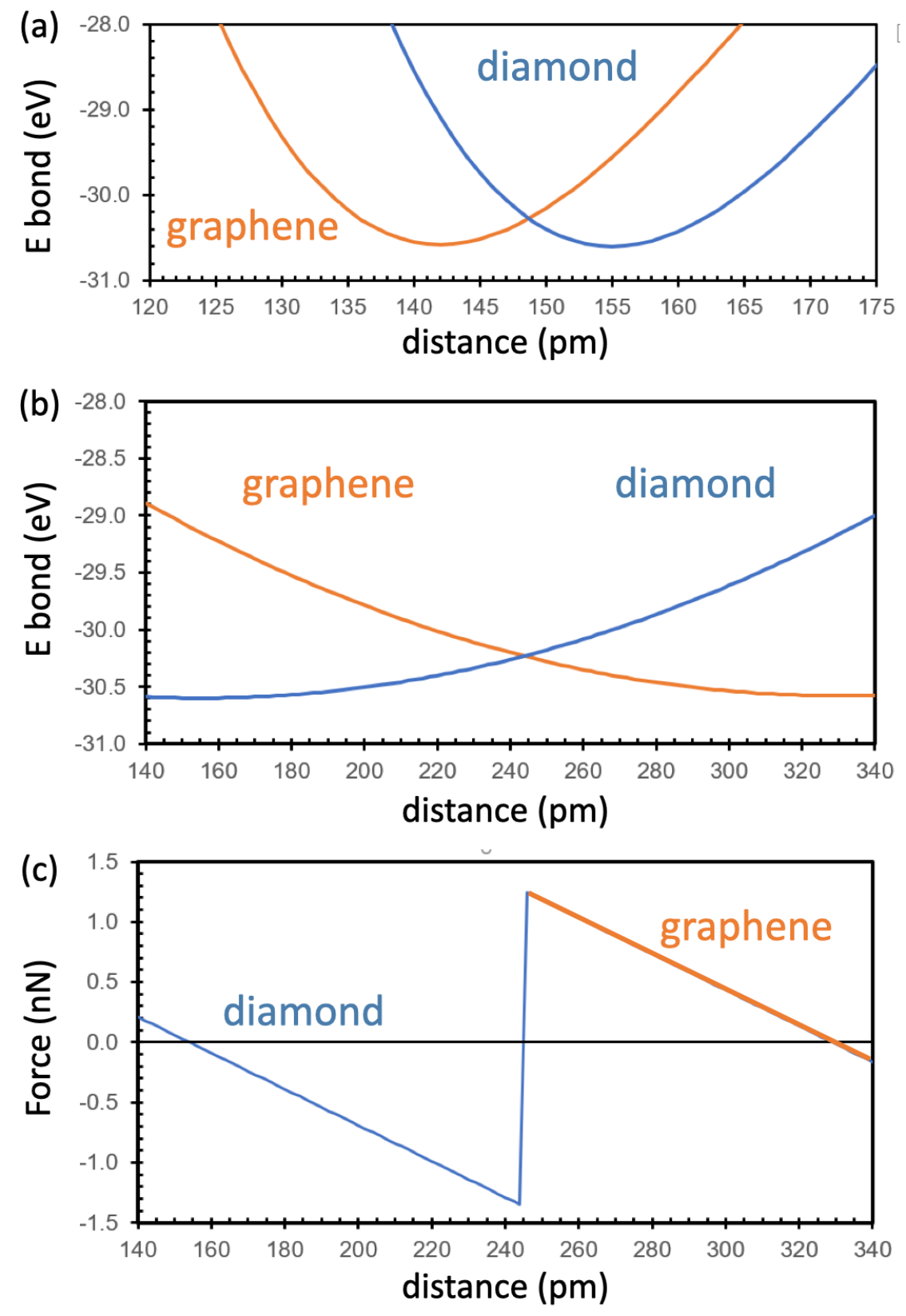}
\end{center}
\caption{(a) Bonding energies of graphene and diamond after Figure 4a of \cite{Popov2019}, showing the energetic minima at the bond lengths of graphene at 142\,pm and diamond at 155\,pm. (b) Energy profile of (a) rescaled to the vertical bond length of bilayer graphene of about 330\,pm. (c) Force in vertical direction derived from (b). Note that (b) and (c) are only qualitative derivations from graphene and diamond data. }\label{fig03_DFT_g_d_trans_struc_F_v_z}
\end{figure}

Figure \ref{fig03_DFT_g_d_trans_struc_F_v_z} (a) displays the bonding energies of graphene and diamond after Figure 4a of \cite{Popov2019}. Note that the bonding energy of graphite and graphene are very similar, the one of graphite is lower by the exfoliation energy. This energy is required to release the van-der-Waals bond between two layers and amounts to approximately 30\,meV per atom. Chemical bonds are typically two orders of magnitude larger. 

The minimum for graphene is located at a distance of 142\,pm, the bonding distance of the sp$^2$ bonds in planar graphene. For diamond, the minimum is at 155\,pm, the bonding distance of the sp$^3$ bonds diamond. 

In bilayer graphene, the layers are spaced by 330\,pm. When rescaling the energy profile of 
fig. \ref{fig03_DFT_g_d_trans_struc_F_v_z} (a) to the bilayer distance, we obtain fig.\ref{fig03_DFT_g_d_trans_struc_F_v_z} (b) and we can take the derivative with respect to distance and obtain 
the force profile in \ref{fig03_DFT_g_d_trans_struc_F_v_z} (c). Note that this procedure is not exact, it is only an approximative method to obtain an idea for a reasonable force versus distance curve in the vertical direction.

\section{Experimental setup}
\label{sec:Experimental}

\subsection{Scanning probe microscopy}
\label{subsec:Scanning probe microscopy}

We use combined STM and AFM to probe the charge density at the Fermi level by STM and the bonding properties of the sample to the various tips by AFM in a single experiment, utilizing a combined scanning tunneling and atomic force microscope operating in ultrahigh vacuum at a temperature of 4.4 K (LT-SPM by ScientaOmicron Nanotechnology, Taunusstein, Germany) employing qPlus force sensors \cite{Giessibl2000APL,Giessibl2019RSI}.

The data with CO tip termination were recorded with a sensor type S1.0 (see table I in \cite{Giessibl2019RSI} for detailed specifications) with a stiffness of $k = 1\,800$\,N/m, an eigenfrequency of $f_0 = 26\,660.3$\,Hz, an amplitude of $A = 50$\,pm and a quality factor of $Q=42\,844$. The bias voltage was 0.1\,V unless noted otherwise.
The data with the CuOx tip were recorded with a sensor type M4 (table I in \cite{Giessibl2019RSI}) with  $k = 1\,850$\,N/m, $f_0 =46\,597.3$\,Hz, $A = 50$\,pm and $Q=482\,321$.  

The instrument is controlled by a Nanonis BP4 SPM electronics with an OC4 oscillation controller (Specs-Nanonis, Zurich, Switzerland) and the higher harmonics were detected with a HF2 Lock-in amplifier (Zurich Instruments, Zurich, Switzerland).

The tip was made of etched tungsten, cleaned by \textit{in situ} field emission and formed by poking into a Cu(111) surface followed by checking the apex of the tip by COFI (carbon monoxide front atom identification). The COFI method for tip characterization uses a CO molecule adsorbed on Cu(111) \cite{Welker2012,Hofmann2014} or Pt(111) \cite{Gretz2020} that images the front section of the tip as illustrated in Fig. \ref{fig21_g_gtip} (c). The COFI method creates a portrait of the tip that shows if a tip has a single atom, a dimer, trimer \cite{Welker2012,Hofmann2014,Gretz2020} or even a flake of graphene as in Fig. \ref{fig21_g_gtip} at the front.

Here, we use single atom metal tips, trimer metal tips, single atom tips terminated by a CO molecule \cite{Bartels1997,Gross2009} and copper oxide (CuOx) tips \cite{Moenig2016,Liebig2019,Yesilpinar2021} that have all been characterized by COFI. Information on the preparation of CO terminated tips and CuOx tips can be found in \cite{Hofmann2014} and \cite{Liebig2019}, respectively.

\subsection{Experimental observables}
\label{subsec:Experimental observables}

We record a set of seven observables in each image in forward and backward scan direction that are used to create images and to analyze the interaction in detail: 
\begin{enumerate}

\item A tunneling current $\langle I\rangle$, where the averaging occurs over many oscillation cycles of the sensor that covers a vertical distance of twice the oscillation amplitude $A$.  
The tunneling current has an exponential distance dependence with gap width $z$ given by
\begin{equation}
I(z)=G_0 V_{bias} \exp(-z\kappa_I)
\label{eq_I(z)}
\end{equation}
where $G_0$ refers to the conductance quantum $G_0=2e^2/ h \approx 1/(12906\,\Omega)$, $e=1.6022\times10^{-19}$\,C is the elementary charge of the electron and $h=6.626\times10^{-34}$\,Js is Planck's constant. The decay constant $\kappa_I$ is given by 
\begin{equation}
\kappa_I=\frac{\sqrt{2m_e \Phi} }{\hbar} 
\label{eq_kappa}
\end{equation}
with electron mass $m_e=9.11\times10^{-31}$\,kg and work function $\Phi$ as explained in \cite{Chen1993}. A typical decay rate of the tunneling current with distance is a reduction to 1/10 for every distance increase of 100\,pm or 1\,\AA{} \cite{Chen1993}.

\item An averaged gradient of the tip sample force is measured by frequency modulation (FM) - AFM \cite{Albrecht1991}. The frequency shift is given by $\Delta f = f-f_0$, where $f$ is the actual oscillation frequency and the frequency shift $\Delta f = f_0 \cdot \langle k_{ts}\rangle/(2k)$ \cite{Giessibl1997PRB}, caused by the tip sample force $F_{ts}$ or more precisely an average over its gradient $k_{ts}=-\partial F_{ts}(z) /\partial z$ as shown in equation \ref{df_F}. 

If the perturbation of the tip-sample force to an oscillating cantilever is weak, the frequency shift is given by
\begin{equation}
\Delta f(z)=\frac{f_{0}}{k\pi}\int_{-1}^1 k_{ts}(z+u A)\sqrt{1-u^2} du
\label{df_F}
\end{equation}
that shows that the average tip sample force gradient $\langle k_{ts}\rangle$ is computed by a convolution of $k_{ts}$ over a semicircular weight function with a radius of the oscillation amplitude $A$ \cite{Giessibl2001APL}.
Experimental noise on this frequency shift measurement can be kept low by experimental precautions and choosing a low acquisition bandwidth \cite{Kobayashi2009,Giessibl2019RSI}.
Note that all distance data $z$ have an arbitrary offset and $z=0$ refers to the minimal distance of the experimental data.

The tip sample force $F_{ts}$ can be recovered from the frequency shift spectrum, using e.g. the Sader-Jarvis formula \cite{SaderJarvis2004} or the Matrix approach \cite{Giessibl2001APL}. Recently, it was discovered that the presence of inflection points in the force curve requires to meet certain conditions regarding the oscillation amplitude $A$ in order to obtain a well-posed force deconvolution \cite{Sader2018,Huber2020,Sader2020}. The deconvolutions performed here have been verified to be well-posed and thus valid.

\item Higher harmonic amplitudes $a_n$ of order $n$ from 2 to 5, where the deflection $q$ of the cantilever is expressed as a Fourier series with $q(t)=\sum_{n=0}^{\infty} a_n \cdot \cos{(n 2\pi f t})$ with $a_1 = A = 50$\,pm and $f = f_0+\Delta f$. These higher harmonics arise when the tip-sample force $F_{ts}$ is not strictly linear with distance $z$. Dürig \cite{Duerig2000} showed that the $n-th$ harmonic $a_n$ can be expressed as a convolution of the force curve with a Chebyshev function and proposed that the full force curve, separated into conservative and dissipative components, can be recovered from the amplitudes and phases of the higher harmonics in FM-AFM. Kawai et al. have used this method to experimentally verify the force field over a KBr(001) surface \cite{Kawai2012}.

\begin{figure}
\begin{center}
\includegraphics[clip=true, width=0.8\textwidth]{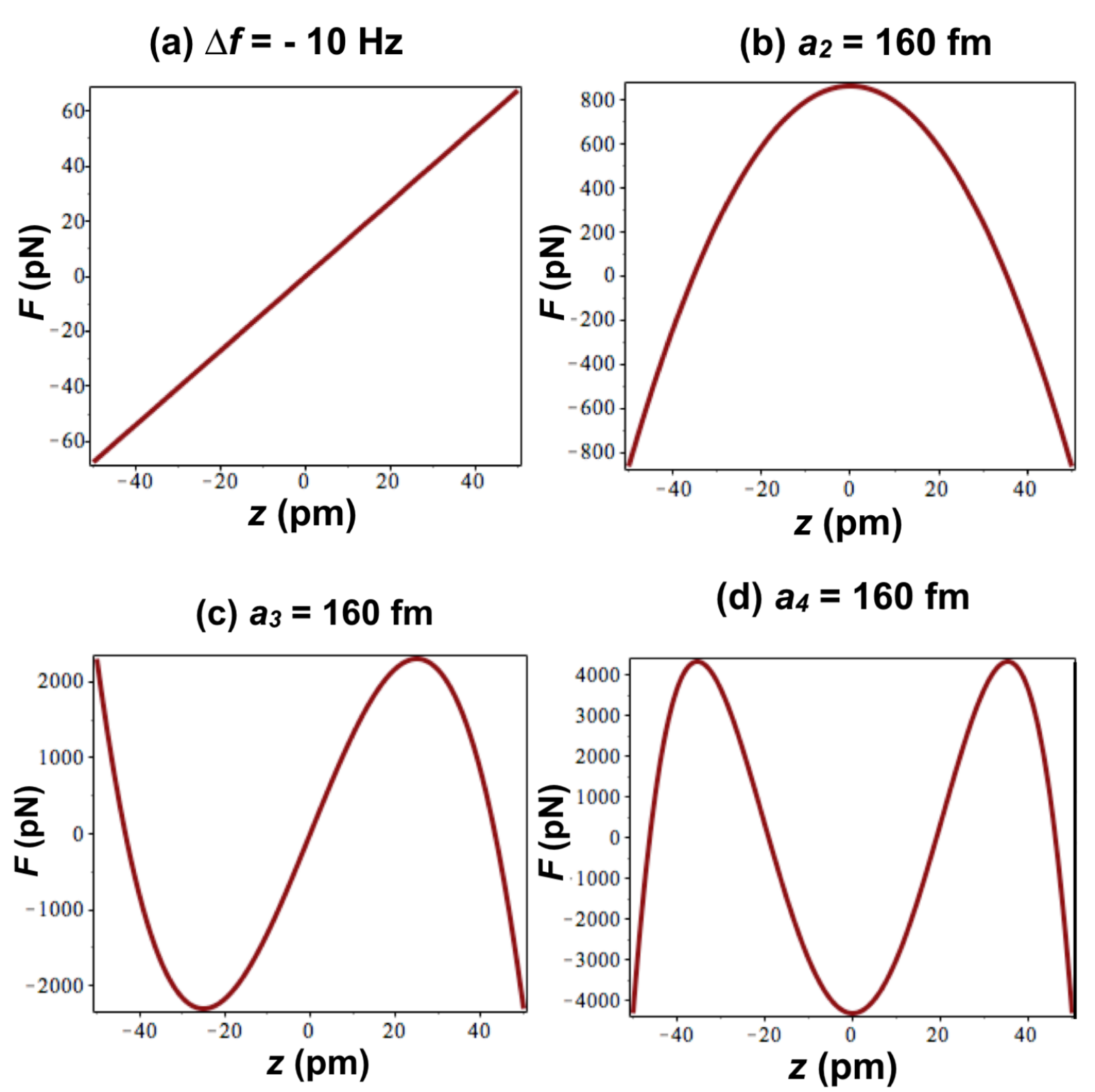}
\end{center}
\caption{Force versus distance curves over a $z-$scale of 100\,pm $=2A$ that are required to (a) generate a frequency shift  of -10\,Hz; (b) a second harmonic $a_2=160$\,fm; (c) a third harmonic $a_3=160$\,fm and (d) a forth harmonic $a_4=160$\,fm. }\label{fig04_df_a2_to_a4}
\end{figure}

The calculation of the higher harmonics as outlined in section VII A.1 of reference \cite{Giessibl2003RMP}, and $a_n$ can be expressed in terms of the Fourier components of the tip-sample force $F_{ts}$ for $n\neq 1$:
\begin{equation}
    a_{n} = \frac{1}{1-n^2}
\frac{1}{\pi (1+\delta_{n0})}
\frac{1}{k}\int_{0}^{2\pi}F_{ts}(z+A \cos(\phi)) \cos(n\phi)  d\phi.
\label{eq_a_n_Fourier}
\end{equation}
The constant cantilever deflection $a_0$ thus is given by the average force divided by the spring constant $a_0=\langle F_{ts}\rangle/k$, the amplitude at the fundamental oscillation frequency is $a_1=A$ and for $n\geq 2$, we find
\begin{equation}
    a_{n} = \frac{1}{1-n^2}\frac{\hat{F}_{ts}(n\omega)}{k}
\label{eq_a_n_Fourier2}
\end{equation}
where $\hat{F}_{ts}(n\omega)$ is the Fourier component of the time dependent tip-sample force 
$F_{ts}(z+A \cos(\omega t))$ at angular frequency $n\omega$. For example, for $F_{ts}(z_0+z)=-1.8\,\textrm{nN}\cdot \cos{(2\arccos{(z/A)})}$ for $-A\leq z \leq A$, we find $a_2=333$\,fm for a sensor stiffness of $k=1800$\,N/m, while for $F_{ts}(z-z_0)=-1.8\,\textrm{nN}\cdot \cos{(3\arccos{(z/A)})}$ for $-A\leq z \leq A$, we find $a_3=125$\,fm. Figure \ref{fig04_df_a2_to_a4} displays force versus distance laws that are needed to generate a frequency shift of $-10$\,Hz and higher harmonics of order 2, 3 and 4 with a magnitude of 160\,fm. It is noted, that measureable frequency shifts require only small force variations, while higher harmonics that rise above the experimental noise level require quite large force variations. 
Mathematically, the $n$-th harmonic $a_n$ can also be evaluated as a convolution of the $n$-th force gradient with bell-shaped weight functions as presented in \cite{Hembacher2004}.

\item A drive signal $V_{exc}$ that excites and maintains a constant oscillation amplitude $A$ of the cantilever. Dissipative tip-sample interactions lead to an increase of the drive signal from $V_{exc}$ to $V'_{exc}$.
The dissipation per oscillation cycle $\Delta E_{ts}$ is given by \cite{Giessibl2019RSI}:
\begin{equation}
\Delta E_{ts} =\frac{\pi kA^2}{Q} (\frac{V_{exc}'}{V_{exc}} -1).
\label{eq_E_diss}
\end{equation}

\end{enumerate}

\subsection{Sample preparation}
\label{subsec:Sample preparation}

To study the graphite to diamond transition induced by an AFM, graphene \cite{Geim2007,Geim2009} on a carbon buffer layer as in graphene grown on the Si side of silicon carbide \cite{Emtsev2009} is an ideal sample, because the graphene top layer is stacked in an AB order \cite{Matsui2015} on top of the rigid SiC substrate.  
Graphene samples were produced by the group of Thomas Seyller by heating silicon carbide in an Argon atmosphere \cite{Emtsev2009}. The samples were inserted into a vacuum chamber, heated in situ for cleaning and transferred into the low temperature scanning probe microscope. This preparation led to a flat sample with large areas of single layer graphene as illustrated in Figure \ref{fig05_g_SiC_overview}.
Two more variations of graphene have been imaged here: a) the surface layer of highly oriented pyrolytic graphite (HOPG) that was prepared by cleaving with the scotch tape cleaving technique and b) monolayer graphene flakes that are known to form from traces of carbon that are naturally dissolved in copper. Heating a Cu (110) crystal to approximately 800\,K leads to migration of carbon to the surface and subsequent graphene formation.

\section{Experimental data}
\label{sec:Experimental Data}

\begin{figure}
\begin{center}
\includegraphics[clip=true, width=0.8\textwidth]{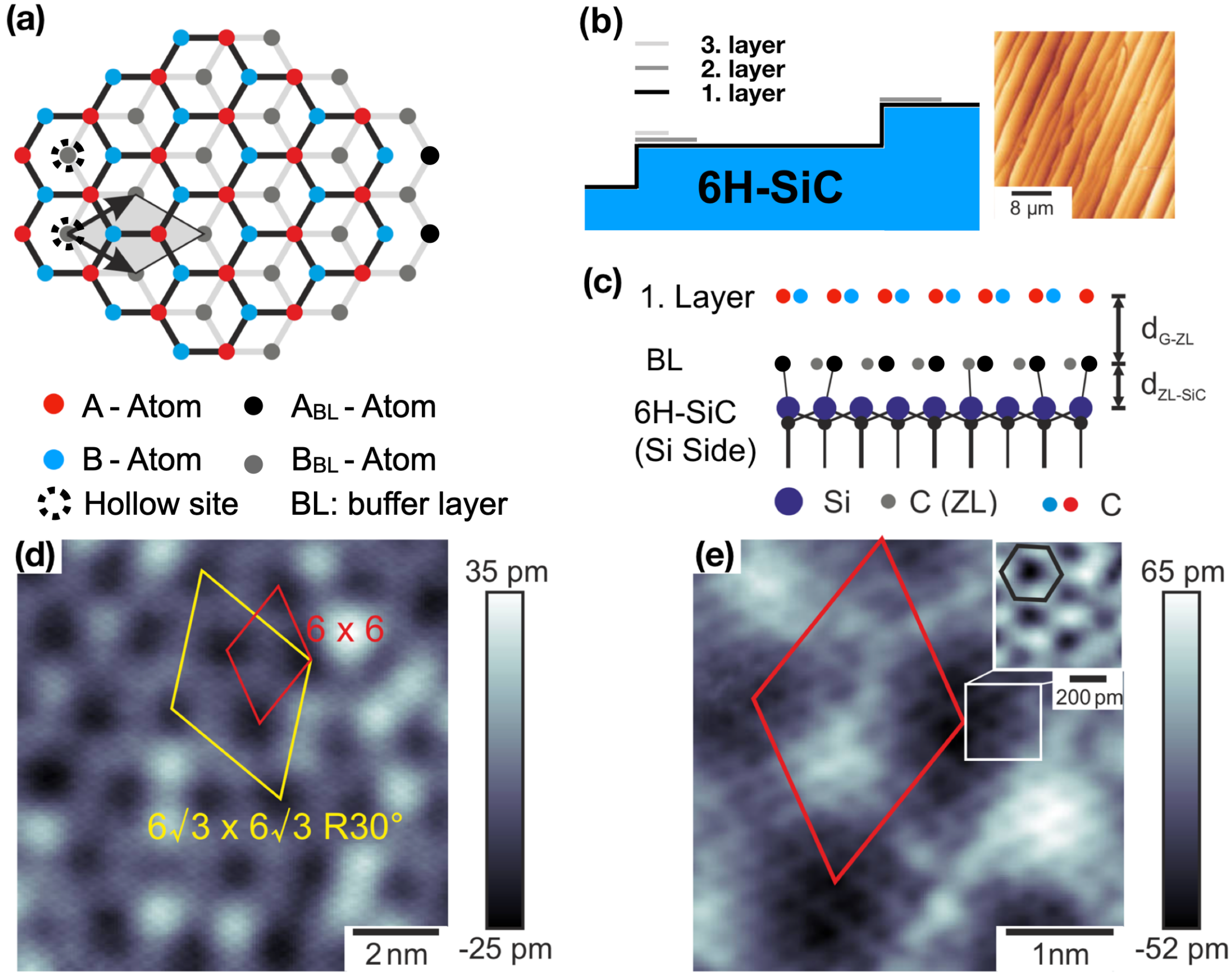}
\end{center}
\caption{(a) Schematic view of monolayer graphene on a buffer layer on top of a SiC crystal. The A sites in the graphene layer are above an atom of the buffer layer, while the B sites are above a hollow site of the buffer layer. (b) Schematic view of SiC substrate with atomic steps, covered by a buffer layer and 1, 2 and 3 layers of graphene. The terraces have a width of a few micrometers as shown in the STM image to the right. (c) Side view of the structure of monolayer graphene on the buffer layer on top of the Si side of SiC. (d) Topographic STM image of the graphene monolayer with its $6\times 6$ and $6\sqrt{3} \times 6\sqrt{3}$ unit cells. (e) Topographic STM image of the graphene monolayer at higher magnification with inset on atomic scale.}\label{fig05_g_SiC_overview}
\end{figure}

Figure \ref{fig05_g_SiC_overview} (a,b,c) shows the structure of monolayer graphene on a buffer layer on top of a SiC crystal. 
The stacking of graphene over the buffer layer leads to two inequivalent lattice sites A and B within the hexagonal unit cell (Fig. \ref{fig05_g_SiC_overview} (a)), where A sits on top of an A$_{BL}$ site on the buffer layer, while B is over a hollow site of the buffer layer (BL).  The B$_{BL}$ site on the buffer layer is underneath a hollow site of the surface graphene layer.

The SiC crystal is atomically flat with terraces with a width of a few $\mu$m separated by atomic steps and covered predominantly by monolayer graphene with patches of bi- and tri-layer graphene (Fig. \ref{fig05_g_SiC_overview} (b)). Topographic STM images in Fig. \ref{fig05_g_SiC_overview}(d,e) show the typical $6\times 6$ domains and $6\sqrt{3} \times 6\sqrt{3}$ unit cells.

\subsection{AFM with inert trimer tips and CO terminated tips}
\label{subsec:AFM with inert trimer tips and CO terminated tips}
\begin{figure}
\begin{center}
\includegraphics[clip=true, width=0.8\textwidth]{fig06_Cu3_CO_tip.pdf}
\end{center}
\caption{Top row: (a) Left: Calculated STM image of graphene for a metal tip (charge density at the Fermi level, arbitrary units). Right: Calculated AFM image for a metal tip (given by total charge density in electrons per nm$^3$, scale inverted). (b) Calculated STM image of graphene for a CO terminated tip (see Fig. \ref{fig26_STM_sim_COtip} for an explanation, arbitrary units). Right: Calculated AFM image of graphene for a CO terminated tip  (given by the total charge density expressed in electrons per nm$^3$). (c)  Experimental constant height STM image (left), simultaneously recorded with the corresponding AFM image on the right recorded with a metal tip. (d) Experimental constant height STM image (left), simultaneously recorded with the corresponding AFM image on the right recorded with a CO terminated tip. Insets show COFI images of these tips, verifying that a metallic trimer tip was used in (c) and a CO-terminated tip was used in (d). Bottom row: Force versus distance curves for the  metallic trimer tip (e) and for the CO terminated tip (f). The insets show how the CO tip termination acts as a spacer, leading to only 1/5 of the vdW force in (f) compared to the trimer metal tip in (e). }\label{fig07_Cu3_CO_tip}
\end{figure}

Figure \ref{fig07_Cu3_CO_tip} shows simulations and experimental STM and AFM data for a metal tip (left two columns) and for a CO terminated tip (right two columns), respectively. The STM contrast is calculated using the Tersoff-Hamann approximation \cite{TersoffHamann1983} for s-type tips in (a) and considering the $p_x/p_y$ symmetry \cite{Chen1990,Gross2011} of the CO terminated tip in (b), see Figure \ref{fig26_STM_sim_COtip}. Carbon atoms appear bright in the STM channel for metal tips (Fig. \ref{fig07_Cu3_CO_tip} (a,c)) and dark for CO tips (Fig. \ref{fig07_Cu3_CO_tip} (b,d)).    

The conductance of undoped graphene is very poor as the density of states at the Fermi level is zero because of the Dirac cone electronic structure. However, the Fermi level is pinned at 400\,meV for epitaxial graphene on the silicon side of silicon carbide such that even for small tunneling bias, a sizeable tunneling current can flow.  

Forces between metal tips and the carbon atoms are attractive \cite{Ondracek2011}, therefore the calculated AFM image in Fig. \ref{fig07_Cu3_CO_tip} (a) and the experimental AFM image in Fig. \ref{fig07_Cu3_CO_tip} (d) show the carbon sites as minima. Figure \ref{fig07_Cu3_CO_tip} (b) is a simulation of STM and AFM data for a CO terminated tip. 
The calculated AFM images use the total charge density as a measure of the attractive force for metal atoms \cite{Ondracek2011} and the repulsive force in case of graphene and CO tips \cite{Boneschanscher2012}.

Figure \ref{fig07_Cu3_CO_tip} (c) shows the experimental STM (left) and AFM (right) images of graphene using a trimer tip with the COFI portrait \cite{Welker2012,Hofmann2014,Emmrich2015} of the tip in the inset of the AFM data. The data shows that the trimer tip is indeed sufficiently inert to allow for atomic imaging in both STM and AFM modes. Although graphite is a lubricant and two graphene layers even show superlubricity \cite{Dienwiebel2004}, single atom metal tips form strong bonds to graphene \cite{Ondracek2011}. The formation of a strong bond between the metal tip and the graphene leads to rehybridization of the sp$^2$ bonds in graphene to sp$^3$ \cite{Perez2022} to unstable imaging and even to ripping the graphene layer off the surface and covering the tip (see experimental data in the appendix in section \ref{subsec:Combined STM and AFM with reactive single atom metal tips}). Just as metal trimers on surfaces are much less reactive than monomers \cite{Berwanger2020}, trimer tips show much less short range attraction than single atom metal tips and thus enable stable imaging of graphene on SiC.
The AFM image appears inverted with respect to the STM image, as the carbon atoms attract the reactive tip and attraction leads to a more negative frequency shift that is displayed dark.

Figure \ref{fig07_Cu3_CO_tip} (d) displays the experimental STM (left) and AFM (right) images of graphene using a CO terminated tip, again with the COFI portrait of the tip in the inset of the AFM data. The STM data shows maxima at the centers of the hexagons as discussed below while the AFM data shows an image of graphene that resembles the total charge density of graphene as expected for a tip that interacts by Pauli repulsion.

Overall, the agreement between calculated data in the top row and experimental data in the middle row of Figure \ref{fig07_Cu3_CO_tip} is very good, deviations can be explained by the unavoidable asymmetries of the trimer tip and the tilt of the CO tip as discussed in the appendix (Fig. \ref{fig26_STM_sim_COtip}).

Figure \ref{fig07_Cu3_CO_tip} (e,f) display the force versus distance curves for the two different tips. Figure \ref{fig07_Cu3_CO_tip} (e) shows a strong attraction between a Cu trimer tip and graphene that reaches 2\,nN where the largest contribution is due to van-der-Waals interaction. The strong attraction between graphene and metal tips is a serious challenge for imaging graphene particularly for single atom metal tips (see Fig. \ref{fig20_sam_tip} in the appendix). Even for trimer tips, it is possible that graphene sticks stronger to the tip than to the buffer layer in SiC graphene. To maintain the integrity of the tip, we did not go close enough to probe the force minimum when recording Fig. \ref{fig07_Cu3_CO_tip} (e). 

Figure \ref{fig07_Cu3_CO_tip} (f) shows a force versus distance curve that is observed for CO terminated tips. The maximal attraction reaches only 200\,pN before repulsion between the CO terminated tip and graphene becomes noticable. The AFM image in Fig. \ref{fig07_Cu3_CO_tip} (d) right displays an experimental image of graphene that looks very similar to the calculated total charge density of graphene as, e.g. shown in Fig. 3D of \cite{Hembacher2003}. The shallow increase of the magnitude of the attractive force with decreasing distance, indicated by blue lines in Figs. \ref{fig07_Cu3_CO_tip} (e,f) points to its van-der-Waals origin. Highly reactive tips that interact by covalent chemical bonding would show an exponential increase of force with decreasing distance $F \propto \exp(-z/\lambda)$ with $\lambda \approx 50$\,pm. Thus, the CO terminated tips that have proven to image organic molecules such as pentacene very well \cite{Gross2009} are also suited well for imaging graphene.

Both force spectra are dominated  by van-der-Waals forces (vdW). The vdW force between a sphere of radius $R$ and a flat surface at distance $z$ is given by \cite{Israelachvili2011}:
\begin{equation}
F_{vdW\,sphere}(z)=-\frac{A_H R}{6 z^2},
\label{eq_vdW_sphere}
\end{equation}
where $A_H$ is the Hamaker constant which has a magnitude of about 1\,eV for many combinations of solids \cite{Israelachvili2011}. A somewhat weaker distance dependence $\propto 1/z$ results for conical and pyramidal tips, where the vdW force for a conical tip with full angle $\alpha$ is given by \cite{Giessibl1997PRB}:
\begin{equation}
F_{vdW\,cone}(z)=-\frac{A_H \tan^2(\alpha/2)}{6 z},
\label{eq_vdW_cone}
\end{equation}

For spherical and conical tips, the vdW force diverges for $z\rightarrow 0$. However, $z$ cannot reach zero as it refers to the distance between the center of the front atom of the tip and the plane connecting the atomic nuclei of the surface layers, yielding a minimal value for $z$ of about 200\,pm. For blunt metal tips, e.g. a spherical tip with a radius of $R=30$\,nm or a conical tip with an opening angle of 170$^{\circ}$, equations \ref{eq_vdW_sphere} and \ref{eq_vdW_cone} yield a maximal vdW attraction of $F_{vdW}(z=200\,\textrm{pm})\approx -20$\,nN for $A_H=1$\,eV.

The force spectra in Figure \ref{fig07_Cu3_CO_tip} (e,f) reveal that the difference in short range force is about 200\,pN between the hollow and the atomic sites for metallic trimer tips in (e), thus we conlude that the 2\,nN attraction is mainly due to van-der-Waals forces. For CO terminated tips, the van-der-Waals forces are much smaller as the CO molecule acts like a separator that keeps the blunt metal tip further away from the flat sample. For the CO terminated tip, the hollow- and atom sites are barely distinguishable in the force spectra for distances of $z>75$\,pm, merging into a difference of about 20\,pN for smaller distances once the Pauli repulsion is in effect.

Experimentally, we can only measure the total tip-sample force $F_{ts}$ by AFM, with 
\begin{equation}
F_{ts}=F_{front\, atom}+F_{background},
\end{equation}
where the background force usually has a long range and is mainly due to vdW forces $F_{vdW}$, unless a large bias voltage is applied that would add electrostatic long range forces.
Hence, it is not possible to separate the long-range van-der-Waals (vdW) part from the short range interactions that originate from the front atom of the tip. When studying individual adatoms, the on-off technique as introduced in \cite{Ternes2011} can be used. In this technique, the short-range force is given by the difference between the tip-sample force spectrum over the adatom and the force spectrum on a flat section of the sample nearby \cite{Ternes2011}. However, this technique is not applicable for flat surfaces, where the vdW contribution has to be guessed from the force characteristics before making contact (blue lines in Figs. \ref{fig07_Cu3_CO_tip} (e,f)). 

\begin{figure}
\begin{center}
\includegraphics[clip=true, width=0.6\textwidth]{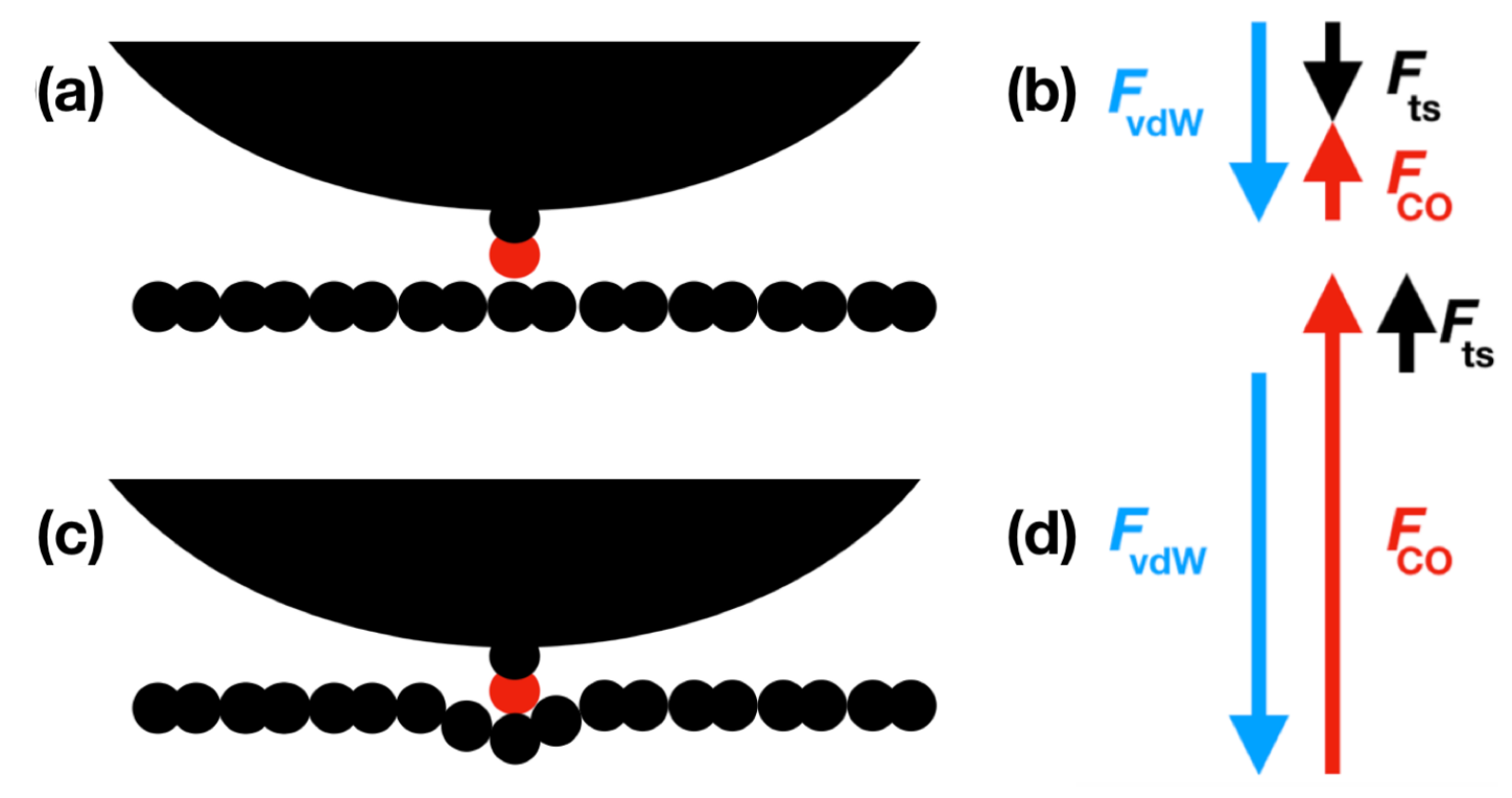}
\end{center}
\caption{Proposed mechanism where the CO tip penetrates into the surface, leading to a very large van-der-Waals attraction $F_{vdW}$ that is compensated by a large repulsive force $F_{CO}$, leading to a very small tip sample force $F_{ts}$.}\label{fig08_FvdW_COpenetration}
\end{figure}

Because of the divergence of the vdW force for small distances, the estimated vdW contribution can be grossly in error for small distances as will be discussed further below. Figure \ref{fig08_FvdW_COpenetration} illustrates where a relatively small total tip-sample force $F_{ts}$ is composed of a strong attractive background force $F_{background}=F_{vdW}$ and a repulsive force of the front atom $F_{front\, atom}=F_{CO}$.

Resolving graphene atomically by STM is straightforward \cite{Geim2007} although the large forces between metallic STM tips and graphene can pull the graphene significantly out of equilibrium, as demonstrated in graphene drumheads \cite{Klimov2012}. Atomically resolved imaging by AFM with metallic tips is more challenging because the atomic spacing of 142\,pm in graphene is only about half of the diameter of a metal tip atom.  Simultaneously recorded STM and AFM images of graphite with tungsten tips have shown large lateral offsets between the current and force channels \cite{Hembacher2003}. This lateral shift has been attributed to the large diameter of metal front atoms and their varying contributions to current and force of different atomic orbitals. For CO terminated tips, where the diameter of the O front atom is only 40\% of the diameter of a metal atom, the registry between STM and AFM images is close to perfect as evident from Figures \ref{fig07_Cu3_CO_tip} (d) and \ref{fig08_CHset_CO_I_df_hh}.

Carbon monoxide terminated tips mounted on qPlus force sensors \cite{Giessibl2019RSI} are now a standard in AFM \cite{note_COtips} after they have successfully been employed to image a pentacene molecule \cite{Gross2009}, other organic molecules \cite{Gross2010,Gross2012}, graphene \cite{Boneschanscher2012} and even molecules that extend into the third dimension normal to the surface \cite{Pavlicek2012}. Here, we find that while CO terminated probes truthfully image graphene, CO bending strongly affects the imaging process on the picometer scale.

\subsection{Constant height data sets recorded with a CO terminated tip}
\label{subsec:Constant height data sets recorded with a CO terminated tip}

\begin{figure}
\begin{center}
\includegraphics[clip=true, width=0.5\textwidth]{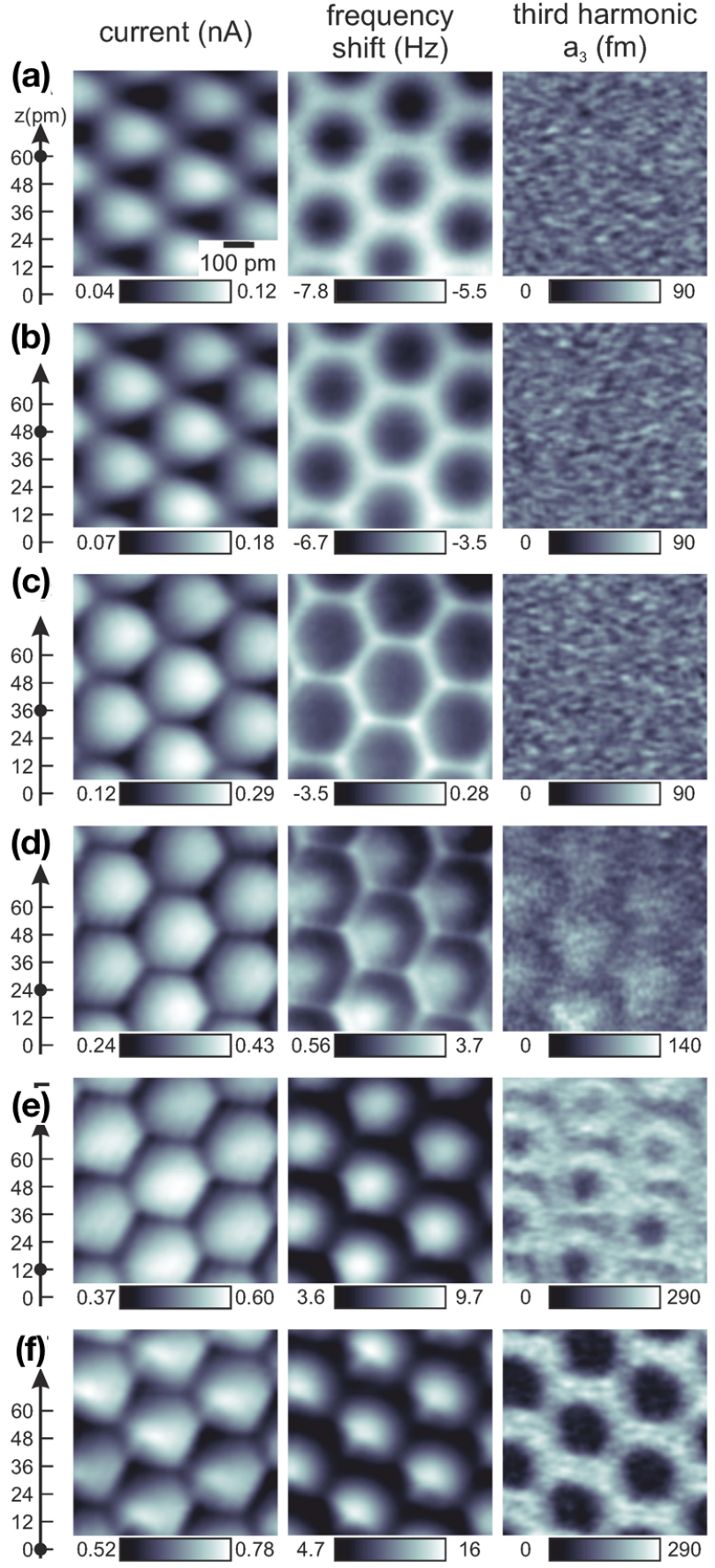}
\end{center}
\caption{Set of constant-height images with tunneling current (left column), frequency shift (center) and third harmonic amplitude $a_3$ (right) over a height interval of 60\,pm spaced by height decrements of 12\,pm from (a) to (f). The tip bias voltage is 100\,mV. The tunneling current is maximal over the hollow sites for all distances, caused by the $p_x/p_y$ symmetry of the CO tip molecule (see Fig. \ref{fig26_STM_sim_COtip}). }\label{fig08_CHset_CO_I_df_hh}
\end{figure}

Figure \ref{fig08_CHset_CO_I_df_hh} shows the detailed contrast development in a set of six constant height images of tunneling current, frequency shift $\Delta f$ and third harmonic amplitude $a_3$. Distances get consecutively closer by decrements of 12\,pm over a range of 60\,pm, starting at (a) the largest distance of $z = 60$\,pm, at (b) $z = 48$\,pm etc.. The left column shows constant height images of the tunneling current, the center column is the frequency shift and the right column shows the third harmonic amplitude. The contrast in the tunneling current (left column) does not change significantly over the small distance variation of 60\,pm from (a) to (f). The current maximum is located at the hollow site for all distances as predicted by the calculation presented in Fig. \ref{fig07_Cu3_CO_tip} (b) left and Fig. \ref{fig26_STM_sim_COtip}. The frequency shift produces an image of graphene similar to those obtained by transmission electron microscopy in Fig. 1 of \cite{Huang2011} or as expected from the total charge density calculation in Fig. 1B at $z = 60$\,pm. For smaller distances, the bonds appear to become sharper from Fig. \ref{fig08_CHset_CO_I_df_hh} (a) center down to Fig. \ref{fig08_CHset_CO_I_df_hh} (c)  center at $z = 36$\,pm. This sharpening effect has been observed in imaging organic molecules and explained already in 2012 \cite{Gross2012}, it is caused by CO bending. 

At $z = 24$\,pm, the contrast starts to invert and the third harmonic $a_3$ rises over its noise level at the hollow sites. Interestingly, of the six monitored higher harmonic amplitudes  $a_2 - a_7$ \cite{Hembacher2004}, only $a_3$ rises above the noise level of about 100\,fm (see Appendix with Fig. \ref{fig23_hh_2to5}). At $z = 12$\,pm, the contrast inversion is fully developed and the higher harmonics show a ring-like structure. The contrast inversion in the frequency shift as well as the rise of the higher harmonics starting at the hollow sites is explained in the following. When the tip is above a hollow site, the oxygen end of the CO molecule becomes trapped in the hollow site and cannot bend laterally. Our hypothesis, supported by additional considerations further below, is as follows. When the local pressure reaches the value that is required to induce two carbon atoms from the graphene surface layer and the underlying buffer layer to rehybridize from three $sp^2$ and one $p_z$ orbital to four $sp^3$ orbitals, a strong anharmonic force component arises that leads to the rise of $a_3$. The Lock-in amplifier that measures $a_3$ is set to a fairly low bandwidth (here, 5\,Hz) to obtain a good signal-to-noise ratio \cite{note_hh_measurement}.  

The frequency shift is more positive over the carbon atoms from frame (a) to (c), pointing to repulsive interactions while a contrast inversion starts to develop at (d) and is fully developed at (e) and (f). The contrast inversion is due to a lateral bending of the CO molecule. The third column, the third harmonic amplitude, becomes noticeable at the height where contrast inversion starts to emerge, explained in section \ref{sec:Density functional theory and proposed generation of higher harmonics}. The apparent bond sharpening that becomes noticeable from (c) in the frequency shift channel is caused by CO bending as discussed in Fig. \ref{fig22_bending_correction} and pertaining text.

\subsection{Interpretation of the constant height data}
\label{sec:Interpretation of the constant height data}

\subsubsection{Calculation of STM data for graphene with CO tips}
\label{subsec:Calculation of STM data for graphene with CO tips}

\begin{figure}
\begin{center}
\includegraphics[clip=true, width=0.6\textwidth]{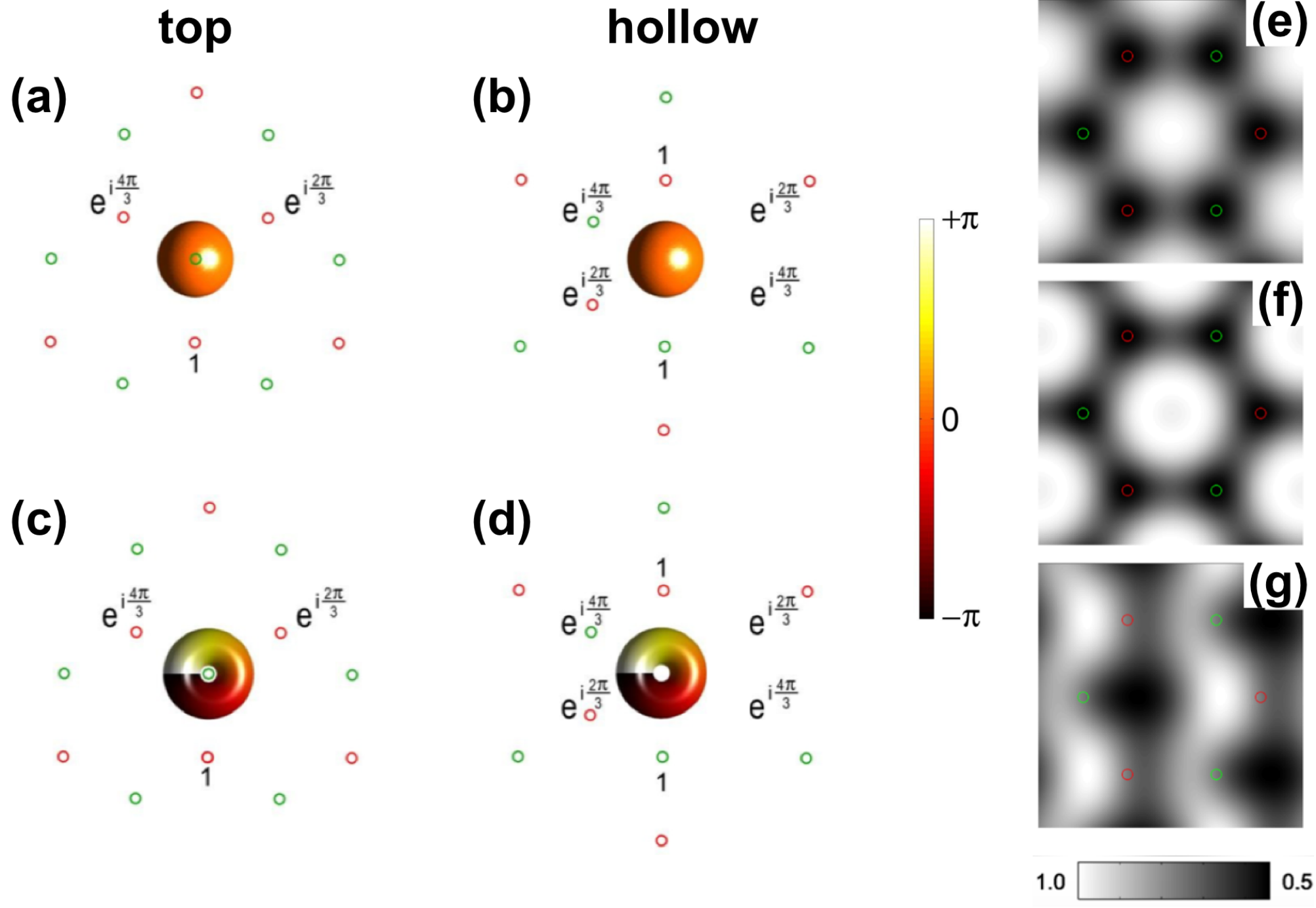}
\end{center}
\caption{(a-d) Symmetry of the graphene Bloch states and the tip wave functions and phases for $s$-tips, $p$-tips ($p$ is short for $p_x/p_y$ here) for hollow and top positions. (f) Simulated constant-height STM image for a CO tip that bends laterally due to lateral forces as it occurs for small distances (see Fig. \ref{fig07_Cu3_CO_tip}(d) left). (g) Simulated constant-height STM image for a CO tip that is slightly tilted by 8$^{\circ}$ in the horizontal direction to match the experimental data of Fig. \ref{fig07_Cu3_CO_tip}.}\label{fig26_STM_sim_COtip}
\end{figure}

The calculation of the STM current for CO terminated tips with its, in comparison to metal tips, inverted contrast as well as the influence of a tilt of the CO tip out of an exact vertical orientation is addressed in Figure \ref{fig26_STM_sim_COtip}. In graphene, electron energies rise linearly with momentum at the Fermi energy rather than quadratic as in many metals and semiconductors \cite{Wallace1947}. The energy-momentum relation is then a Dirac cone, where the zero in energy corresponds to the Fermi wave vector $k_F$ at the K-points. 
The STM current has been calculated assuming that the $\pi^*$ orbital of the CO tip dominates the symmetry of the tip wave function. The derivative rule by Chen \cite{Chen1990} yields a current proportional to the modulus of the gradient of the local density of states, as it was experimentally verified in single molecules \cite{Gross2011}. The local density of states is calculated for graphene within a tight binding model, taking into account only $p_z$ orbitals. The tilting of the molecule ensures contributions to the current proportional to the $z$-derivative, absent for a CO molecule standing perfectly perpendicular to the surface. The tilting angle has been taken as a fitting parameter.

Figure \ref{fig26_STM_sim_COtip} (a-d) shows the Bloch states at the Fermi energy with Fermi wave vector at the K-points of the Fermi surface $\vec{k}_F = \frac{1}{3} (\vec{b}_1 - \vec{b}_2$), where $\vec{b}_1$ and $\vec{b}_2$ denote the reciprocal unit vectors in $\vec{k}$-space \cite{Wallace1947}. For simplicity we have considered undoped graphene and the Bloch states associated to the Fermi wave vector $\vec{k}_F = 1/3 \vec{b}_1 - 1/3 \vec{b}_2$, where $\vec{b}_1$ and $\vec{b}_2$ denote the reciprocal unit vectors in $\vec{k}$-space. (e) Simulated constant-height STM image for a CO tip ($p$-tip). Thus, we find a maximal current in the center of the graphene hexagons and zero current at the atom positions as displayed in Figure \ref{fig26_STM_sim_COtip} (e). The effect of the tip-sample forces on tilting the CO molcule are shown in Figure \ref{fig26_STM_sim_COtip} (f), and an 8 degree tilt from a vertical alignment towards the $y-$direction and the p-wave CO tip is reflected in the calculated image in Figure \ref{fig26_STM_sim_COtip} (g).

\subsubsection{Simulating frequency shift data with the probe particle model}
The interpretation of AFM data on organic materials recorded with CO terminated tips is often performed using the probe-particle model after Hapala et al. \cite{Hapala2014}. This model assumes a Lennard-Jones interaction between the O front atom of tip and allows for a lateral bending of  the CO termination. Standard parameters after \cite{Hapala2014} were used to plot $\Delta f$ images in steps of 12 pm. For the simulations (as in the experiment), $k = 1800$ N/m and $f_0 = 30300$ Hz. The graphene is a flat sheet of carbon that was not allowed to relax. No electrostatics were included in the calculation.

\begin{figure}
\begin{center}
\includegraphics[clip=true, width=0.6\textwidth]{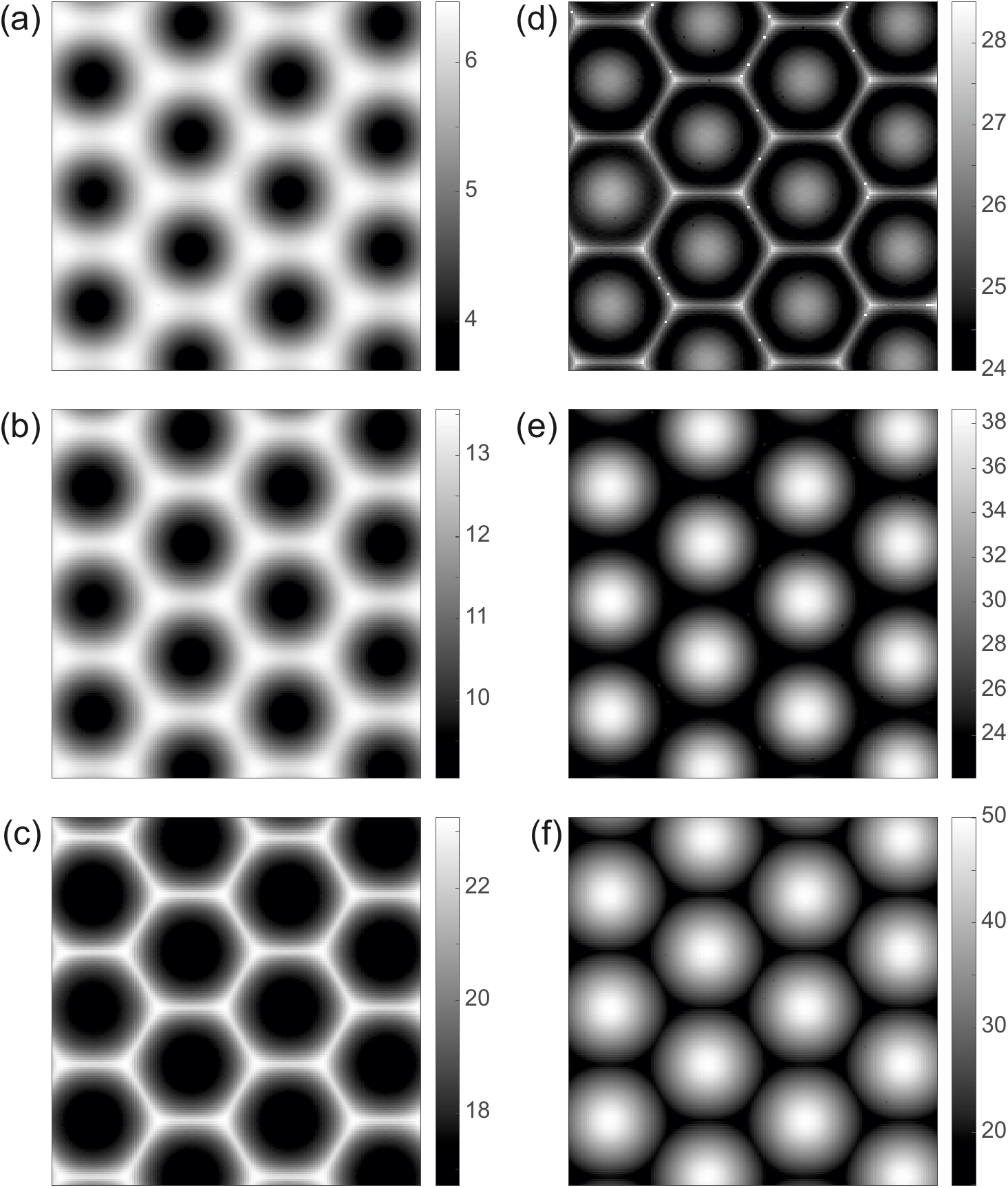}
\end{center}
\caption{Effect of CO bending on image contrast and force spectroscopy. 
(a-c) Constant height AFM images of the average tip-sample force gradient $\langle k_{ts}\rangle=2 k \Delta f / f_0$ recorded at three distances where the $z\rq{}$ scale corresponds to the $z-$scale in Figure \ref{fig08_CHset_CO_I_df_hh}. 
(d) Contour line at larger distance where little CO bending occurs, yielding regular contrast where atoms appear bright.
(e) Contour line at close distance, where the O termination of the CO tip remains in the center of the graphene hexagon all the time, leading to strong CO bending and contrast inversion where hollow sites appear bright.
(f) Force versus distance spectra in the center of the hexagon or hollow site (open circles) and over an atom position (solid diamonds), showing a kink followed by a smaller slope at a distance of $\approx 40$\,pm caused by CO bending.  }\label{fig10_PPM}
\end{figure}

A comparison of the experimental frequency shift data in Figure \ref{fig08_CHset_CO_I_df_hh} with the probe-particle model  simulations in Fig. \ref{fig10_PPM} shows a very nice agreement, starting with an image that resembles the total charge density of graphene in (a), apparent bond sharpening from (b-d). This apparent bond sharpening with decreasing distances was already observed by Gross et al. in 2012 in Fig. 2 of \cite{Gross2012}. Further distance reduction leads to incipient and complete contrast reversal from (d) to (f). CO bending also is responsible for a crossover in local force versus distance spectra.

\subsubsection{CO bending leads to crossover in force versus distance spectra}
When the CO terminated tip is directly above an atom and interacts via Pauli repulsion, it is in an unstable equilibrium. As the lateral stiffness of CO terminated tips is small, only about 0.25\,N/m \cite{Weymouth2014}, small lateral forces will bend the CO tip termination to the side.

\begin{figure}
\begin{center}
\includegraphics[clip=true, width=0.6\textwidth]{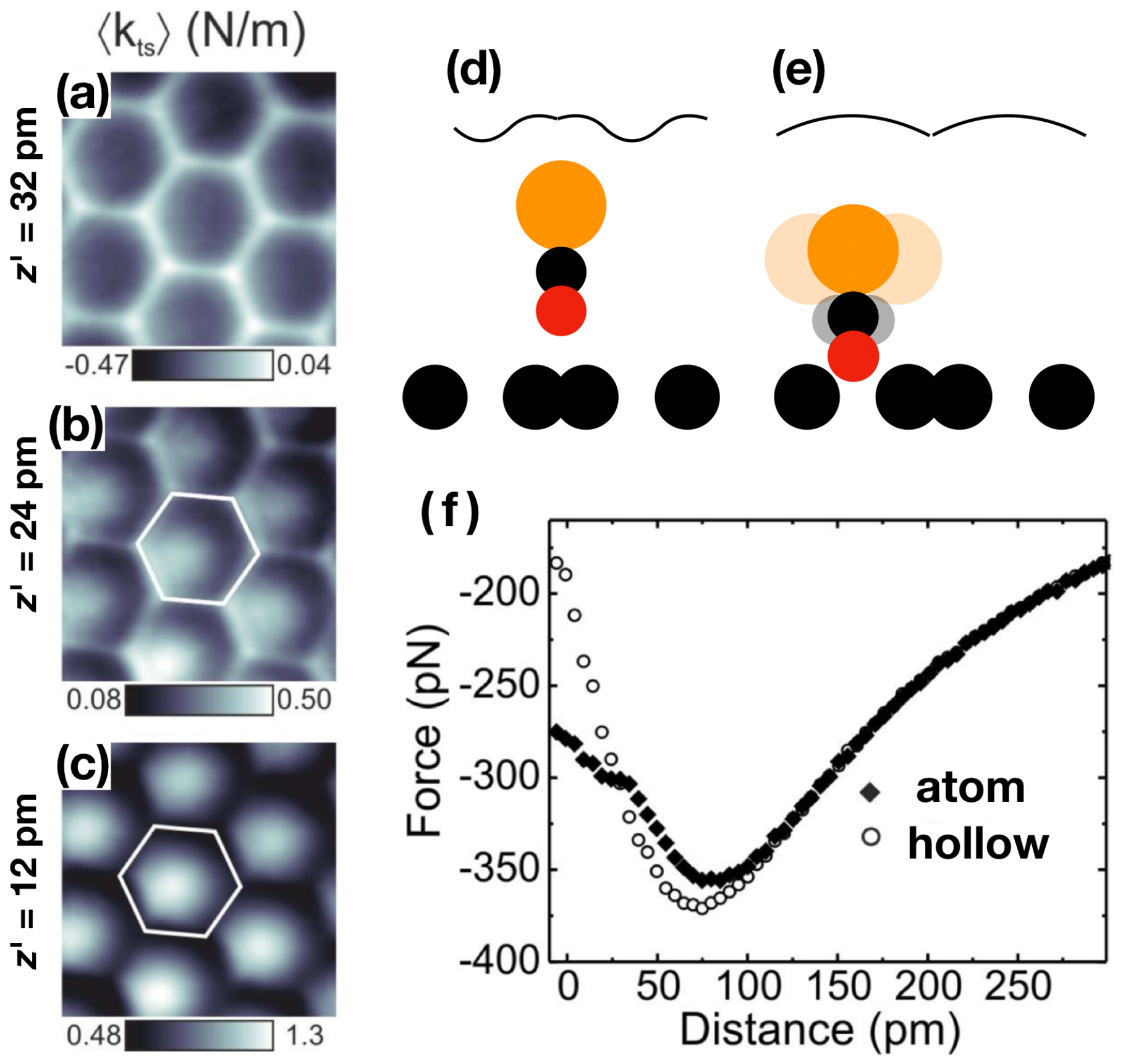}
\end{center}
\caption{Effect of CO bending on image contrast and force spectroscopy. 
(a-c) Constant height AFM images of the average tip-sample force gradient $\langle k_{ts}\rangle=2 k \Delta f / f_0$ recorded at three distances where the $z\rq{}$ scale corresponds to the $z-$scale in Figure \ref{fig08_CHset_CO_I_df_hh}. 
(d) Contour line at larger distance where little CO bending occurs, yielding regular contrast where atoms appear bright.
(e) Contour line at close distance, where the O termination of the CO tip remains in the center of the graphene hexagon all the time, leading to strong CO bending and contrast inversion where hollow sites appear bright.
(f) Force versus distance spectra in the center of the hexagon or hollow site (open circles) and over an atom position (solid diamonds), showing a kink followed by a smaller slope at a distance of $\approx 40$\,pm caused by CO bending.  }\label{fig11_CObend_spec}
\end{figure}

Figure \ref{fig11_CObend_spec} shows two consequences of CO bending. 
First, we again observe apparent bond sharpening in Fig. \ref{fig11_CObend_spec} (a), transiting into incipient (b) and full (c) contrast reversal. The mechanism for this is illustrated in Fig. \ref{fig11_CObend_spec} (d,e). For intermediate distances, the CO tip termination stays almost perfectly upright and the contour of the frequency shift at constant height would show a positive value right above the carbon atoms as sketched in Fig. \ref{fig11_CObend_spec} (d). When moving very close, the oxygen termination of the CO tip is caught in the hollow site of the graphene hexagon and the CO termination pivots around the hollow site of graphene, causing a contrast inversion in the constant height frequency shift images in Figs. \ref{fig11_CObend_spec} (a-c). 

The second consequence of CO bending is apparent in the force versus distance spectra shown in 
Figure \ref{fig11_CObend_spec} (f). The force versus distance spectrum over the center of the hexagon results in a smooth force distance curve because the oxygen end of the CO terminated tip remains stable in the center of the hexagon while local repulsive forces start to act. Note that although the net force on the tip is still attractive at a distance of 100\,pm, the force on the CO tip is already repulsive, i.e. positive. The strong vdW attraction of the tip that is larger in magnitude than the repulsion to the CO tip leads to a net negative force on the tip. When performing the force spectrum over the atom position in Figure \ref{fig11_CObend_spec} (f), the CO molecule at the tip starts to bend at a distance of around 40\,pm, leading to a crossover of the two force curves. 

An inverted crossover, caused by a completely different reason, has been observed by Boneschanscher et al. in Fig. 1c of \cite{Boneschanscher2012} for graphene on iridium using a metallic tip. The crossover region in Fig. 1c of \cite{Boneschanscher2012} is smooth and was explained by turning an initially attractive short range interaction between the tip and the carbon atoms in graphene to repulsion for smaller distances. COFI images of the tip are not presented in \cite{Boneschanscher2012}, and as these authors did not report instability issues as those observed here in Figure \ref{fig20_sam_tip}, we suspect that either graphene adheres much stronger to Ir(111) than to the buffer layer on SiC, or their tip was also a fairly inert metallic trimer tip similar to the one used in Fig. \ref{fig07_Cu3_CO_tip} rather than a highly reactive single atom metal tip as presented in Fig. \ref{fig20_sam_tip}.

\subsubsection{The emergence of higher harmonics for close distances}
The right column of Figure \ref{fig08_CHset_CO_I_df_hh} shows that the third harmonic $a_3$ starts to become noticeable at a distance (row (d)) where the contrast inversion starts to emerge and the repulsive force between the CO tip termination becomes large enough to cause considerable tip bending.

The emergence of a third harmonic requires the tip to go through a force profile as plotted in Figure \ref{fig04_df_a2_to_a4} (c), i.e. overcoming a force barrier, followed by a reduction in force and again followed by an increasing force. The qualitative Figure \ref{fig03_DFT_g_d_trans_struc_F_v_z} (c) would provide a similar force profile, suggesting that the emergence of $a_3$ is related to overcoming a strong force barrier such as occurs in the graphite to diamond transition. 

\begin{figure}
\begin{center}
\includegraphics[clip=true, width=0.8\textwidth]{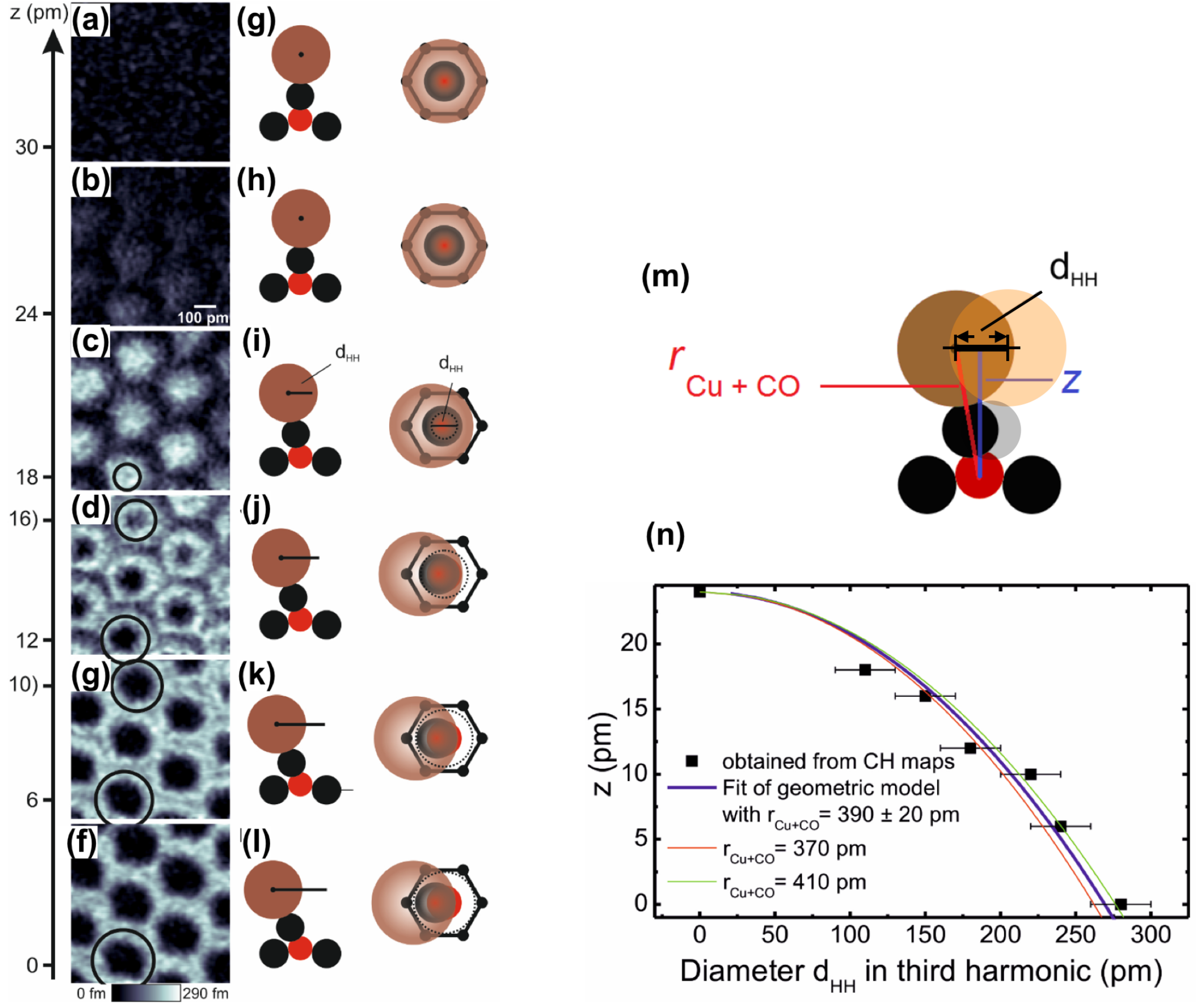}
\end{center}
\caption{(a) Third harmonic amplitude $a_3$ for six different heights spaced by 6\,pm. A noticeable amplitude $a_3$ first occurs in the center of the graphene hexagons in (b), turning to rings for closer distances in (d) with increasing diameters $d_{\textrm{HH}}$ (g,f). Note that the scan frame is not perfectly parallel to the sample surface and the $z$-distances at the lower edge are approximately 4\,pm smaller than at the top edge. (g-l) Schematic view showing a tilted CO tip termination. (m,n) The ring diameter $d_{\textrm{HH}}$ follows a simple Pythagorean relation $(d_{\textrm{HH}}/2)^2 = r_{Cu+CO}^2-z\rq{}^2$. }\label{fig12_hh_ring_dia}
\end{figure}

While $a_3$ starts to emerge at the hollow sites, further approach leads to a ring-like appearance. 
We now look closer at this emergence of rings in $a_3$ in Figure \ref{fig12_hh_ring_dia}. The weak lateral stiffness of the CO terminated tip allows it to pivot around the center of the metallic front atom of the tip with an arm of $d=r_{Cu}+r_{CO}$ as displayed in Fig. \ref{fig12_hh_ring_dia}. 

Figure \ref{fig12_hh_ring_dia} (a) shows the evolution of third harmonics $a_3$ with decreasing distance at distance decrements of merely 6\,pm. Note that there is a slight tilt of about 0.5$^\circ$ in the $500\,\textrm{pm}\times 500\,\textrm{pm}$ scan frames - the top line is approximately 4\,pm closer than the bottom line, therefore $a_3$ becomes noticeable first at the bottom of Figure \ref{fig12_hh_ring_dia} (b). The third harmonics start to emerge at the hollow sites and evolve into ring-like structures for closer distances, in accordance with the model of their origin presented in Figure \ref{fig12_hh_ring_dia} (m) that shows that the diameters of the ring-like structures follow a simple geometric relation $(d_{\textrm{HH}}/2)^2 = r_{Cu+CO}^2-z\rq{}^2$ plotted in Fig. \ref{fig12_hh_ring_dia} (n).

\begin{figure}
\begin{center}
\includegraphics[clip=true, width=0.6\textwidth]{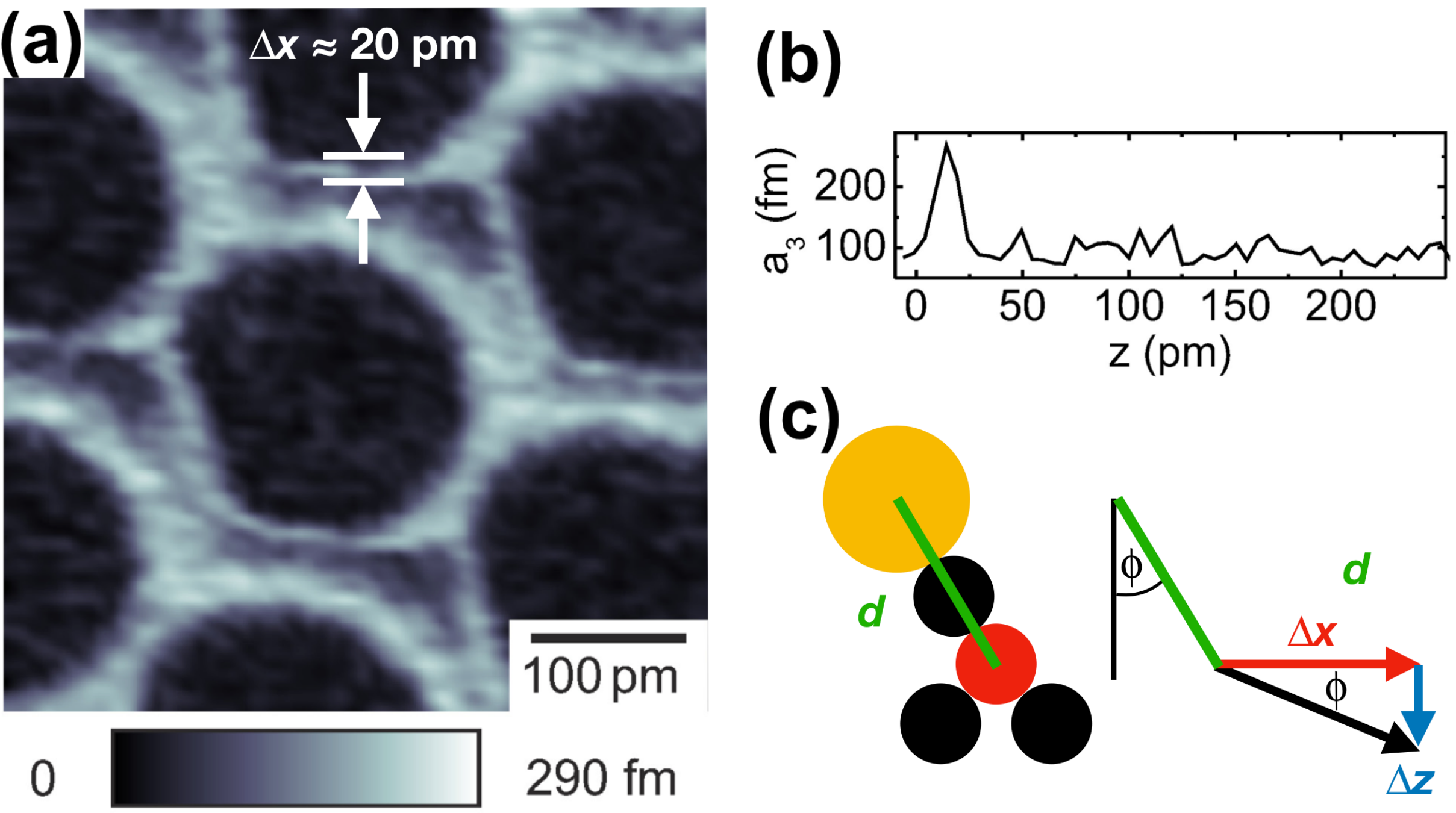}
\end{center}
\caption{(a) High resolution image of $a_3$ close to the distance where two adjacent higher-harmonic rings merge. Parameters: $V_{tip} = 0$\,V, image size $0.5\times 0.5$\,nm$^2$ at   $64\times 64$\,pxl, scanning speed 125\,pm/s. (b) Distance spectrum of the third harmonic amplitude $a_3$ recorded over the hollow site. (c) Geometry of the CO terminated tip trapped with its O termination in the hollow site (see text).}\label{fig13_a3_gen_width_ch}
\end{figure}

Figure \ref{fig13_a3_gen_width_ch} (a) shows a constant height image of the third harmonic $a_3$ at a close distance where the rings of adjacent unit cells merge, leaving a triangular local minimum at one of the two lattice sites A and B. The small width $\Delta x \approx$ 20\,pm of the roughly circular profile of $a_3$ shows that the distance dependence of $a_3$ at the hollow site plotted in Fig. \ref{fig13_a3_gen_width_ch} (b) even overestimates the $z$ width of $a_3$. With an estimate for the arm length $d$, we find a maximal deflection $\phi \approx 18^{\circ}$, and with the geometry considerations in Figure \ref{fig13_a3_gen_width_ch} (c) we find $\Delta z \approx \Delta x \times \tan{\phi} \approx 7$\,pm.

\begin{figure}
\begin{center}
\includegraphics[clip=true, width=0.6\textwidth]{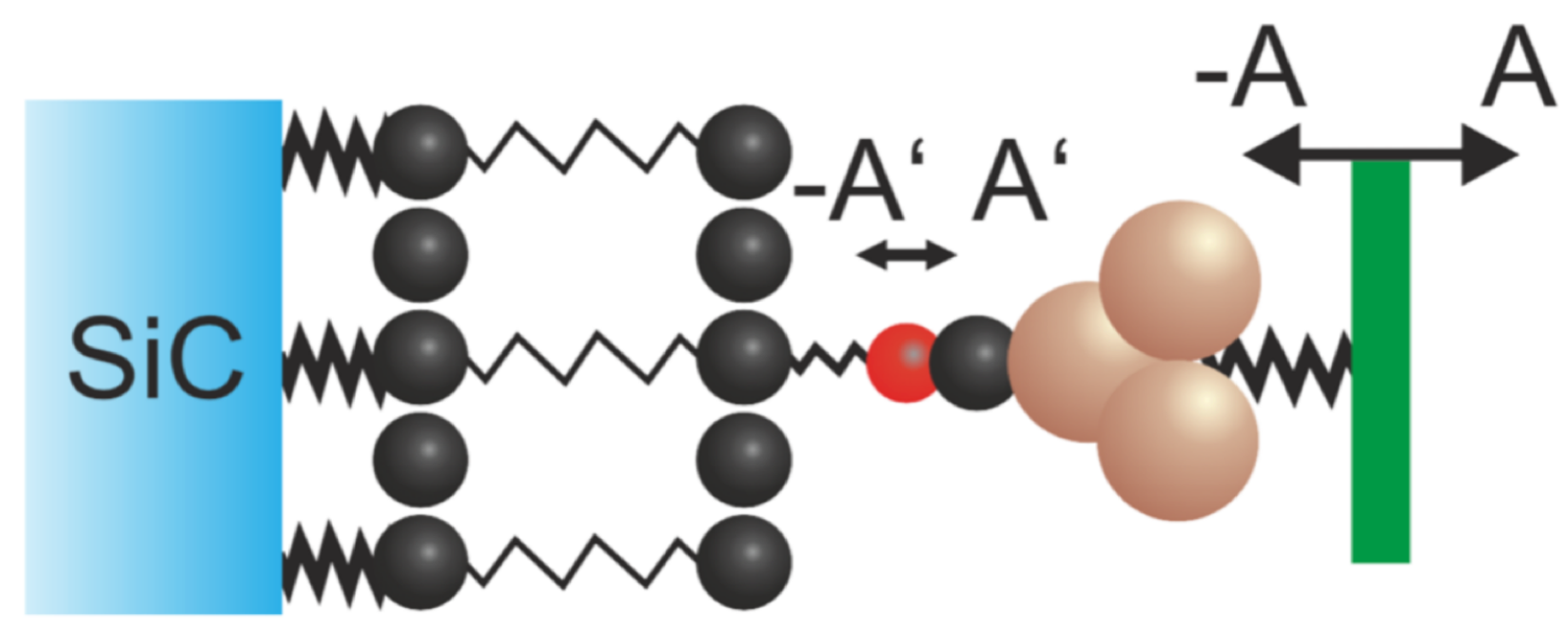}
\end{center}
\caption{Geometry of the SiC substrate, covered by a buffer layer of graphene and an additional graphene layer. The CO terminated tip oscillates at a reduced amplitude $A'$ once the repulsive forces become significant (see text).}\label{fig14_tip_elast}
\end{figure}

Thus, the third harmonic occurs over a very thin slize in $z$ direction of only about 7\,pm. Given that the weight function to derive $a_3$ (see equation \ref{eq_a_n_Fourier}) has a width of about $A'$, it shows that the reduced oscillation amplitude between the tip's front atom and graphene in the squeezed material is only a fraction of $A=50$\,pm. This reduced amplitude can be explained by the repulsive load of the tip over the relatively soft bilayer graphene. Figure \ref{fig14_tip_elast} is a schematic view of the sample and tip arrangement. Although the tip oscillates at amplitude $A$, the distance between the graphene surface layer and the tip will oscillate at a much smaller amplitude $A'$ once the repulsive load becomes large enough to compress the soft bonds between the buffer layer and the graphene.

\begin{figure}
\begin{center}
\includegraphics[clip=true, width=0.3\textwidth]{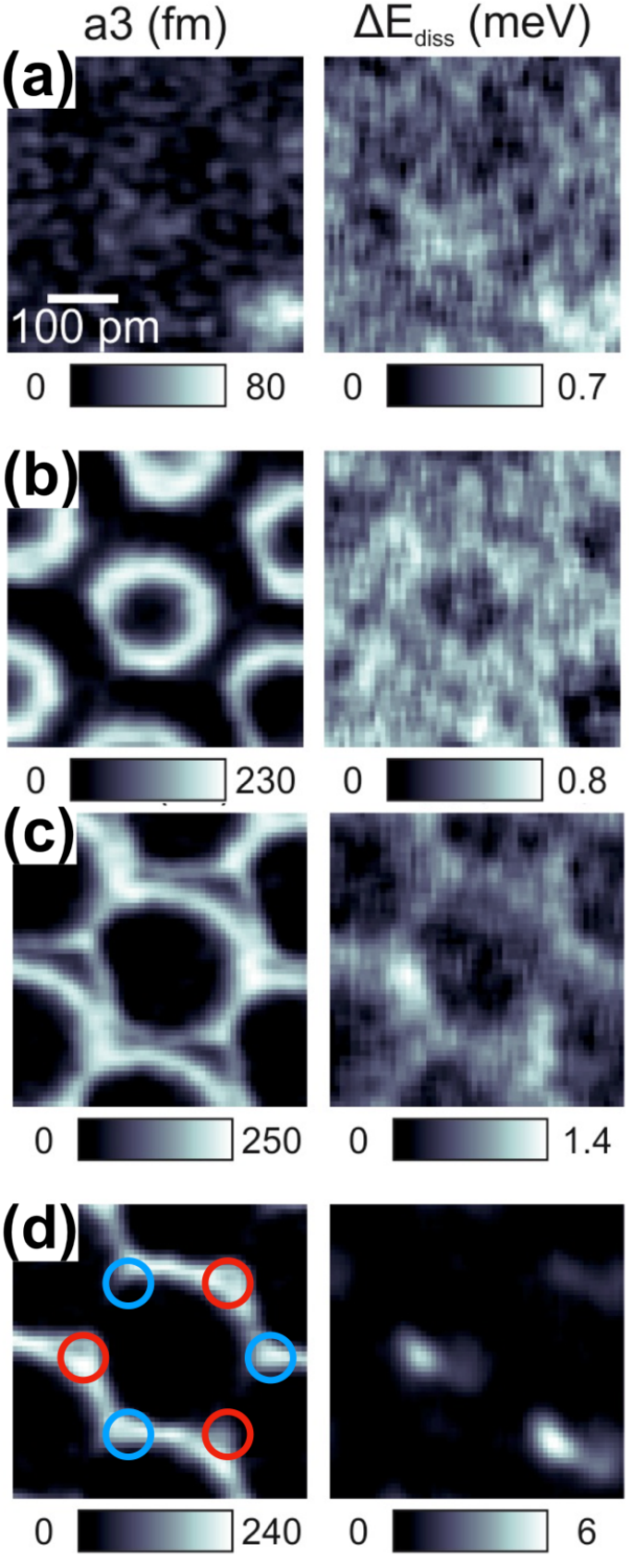}
\end{center}
\caption{(a-d) Third harmonic amplitude $a_3$ (left) and energy dissipation of the sensor per oscillation cycle (right) at distance decrements of merely 5\,pm per frame, i.e. the whole $z-$range covered here is only 20\,pm. The onset of a few meV of dissipation for very small distances indicates a hysteresis in the force versus distance curve. Parameters: $V_{tip} = 0$\,V, image size $500\times 500$\,pm$^2$ at   $64\times 64$\,pxl, scanning speed 125\,pm/s.}\label{fig15_hh_dissipation}
\end{figure}

Figure \ref{fig15_hh_dissipation} shows another high resolution image of $a_3$ supplemented by dissipation data on the right column. The presence of dissipation after equation \ref{eq_E_diss} shows that the tip oscillated non-adiabatically, exciting phonons of a large amplitude as present at elevated temperatures.
 
In a single graphene layer, A and B sites are equal. As mentioned above, for graphene on the buffer layer of SiC, A and B sites are different, because A sites have A$_{BL}$ atoms in the buffer layer underneath while B sites are above hollow sites. If the hypothesis about the graphite to diamond bonding transition is correct, an asysmmetry between A and B sites is expected. We expect the A sites to be reduced in height with respect to the B sites. Figure \ref{fig15_hh_dissipation} shows high resolution data of $a_3$ and the energy dissipation per oscillation cycle for four different heights spaced by merely 5\,pm. The third harmonic $a_3$ shows clear differences between A and B sites in (c) and (d), and the dissipation data shows these differences clearly in (c) and in particular in (d), supporting the bonding transition hypothesis.

\subsection{Distance spectra}
\begin{figure}
\begin{center}
\includegraphics[clip=true, width=0.8\textwidth]{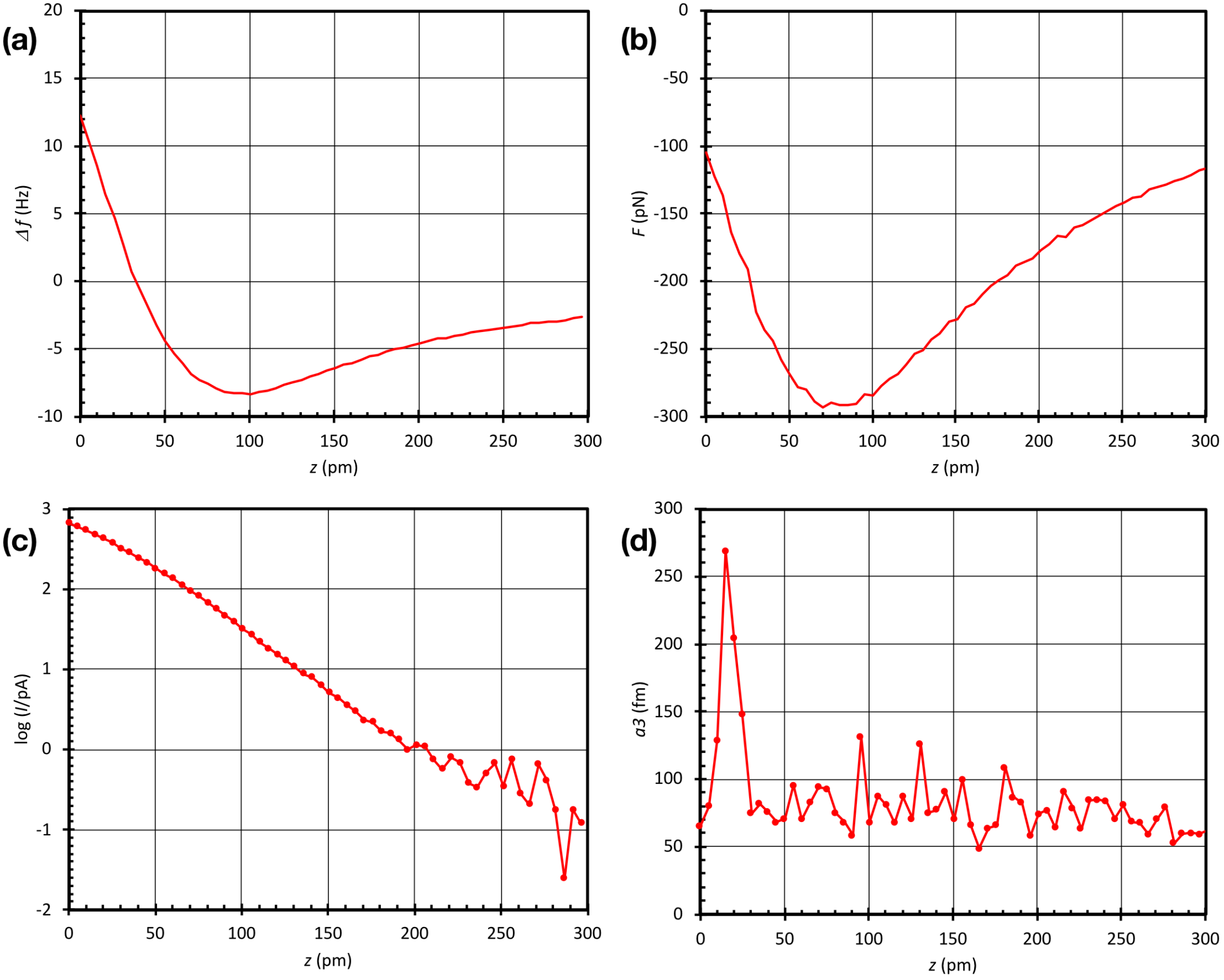}
\end{center}
\caption{Distance dependence of (a) frequency shift $\Delta f(z)$, (b) tip-sample force $F_{ts}(z)$, (c) logarithm of tunneling current in pA and (d) third harmonic amplitude $a_3(z)$ as a function of vertical distance $z$. The data was recorded in the center of a hexagon (hollow site), see Fig. \ref{fig05_g_SiC_overview} (a).}\label{fig16_df_F_I_a3_spectra}
\end{figure}

The physical nature of the tip-sample interaction is revealed by the distance dependence of the pertinent physical observables. Figure \ref{fig16_df_F_I_a3_spectra} (a) shows the distance dependence of the frequency shift recorded with a CO terminated tip that was placed over a hollow site, i.e. the center, of a graphene unit cell. The deconvoluted force in Fig. \ref{fig16_df_F_I_a3_spectra} (b) has only a very small magnitude, reaching merely -300\,pN. However, this inconspicious force curve is incompatible with the distance dependence of two more observables, the averaged tunneling current and the third harmonic $a_3$.

The averaged tunneling current $\langle I \rangle$ is displayed in a logarithmic scale in Fig. \ref{fig16_df_F_I_a3_spectra} (c), it is at or below the noise level of about 1\,pA for $z>200$\,pm, rising in the expected exponential fashion after equation \ref{eq_I(z)} until $z$ reaches about 70\,pm, where the slope becomes noticably smaller. This reduction in slope points to the emergence of repulsive forces, as previously observed in local tunneling versus distance spectroscopy by Schull et al. \cite{Schull2009} on buckyballs.  

The emergence of a strong third harmonic in Fig. \ref{fig16_df_F_I_a3_spectra} (d) is interesting and possibly caused by the graphite to diamond transition discussed earlier. The force versus distance relation that is needed to generate a third harmonic of a magnitude of 0.2\,pm, again derived from equation \ref{eq_a_n_Fourier}, is a force that swings up and down by almost 3\,nN as displayed in Figure \ref{fig04_df_a2_to_a4}.

\section{Images of graphene using CuOx tips}

The introduction of copper oxide (CuOx) tips \cite{Moenig2016} provided tips that are as inert as CO terminated tips but more stiff.  

The termination of AFM tips with a CO molecule has three profound benefits: First, CO terminated tips are chemically highly inert, interacting mainly by Pauli repulsion \cite{Moll2010}, where so far, our group only found Cu and Fe as exceptions \cite{Huber2019}. Second, the atomic radius of O is small. Therefore, they allow high spatial resolution. The lateral softness can lead to imaging artifacts, although sometimes the flexibility of the tips provides greater contrast.
Third, the long CO molecule acts like a distance holder that keeps the blunt metal tip further from the flat surface, reducing vdW attraction.

In 2016, Mönig et al. \cite{Moenig2016} introduced copper oxide (CuOx) tips, tips that share the first and second benefit of CO tips yet providing greater lateral stability. We have used those tips to image graphene on SiC as well.

\begin{figure}
\begin{center}
\includegraphics[clip=true, width=0.5\textwidth]{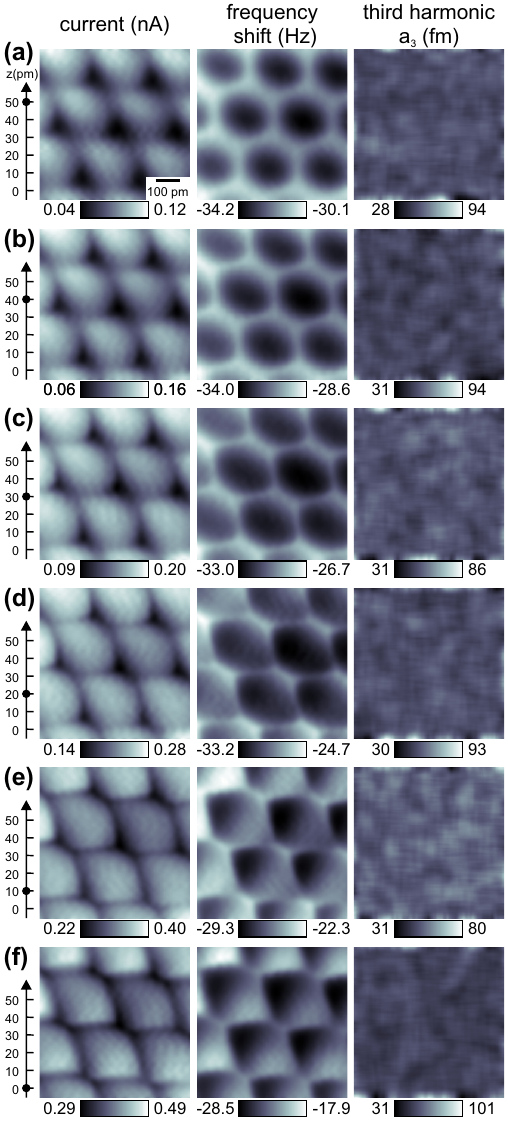}
\end{center}
\caption{Constant-height images of tunneling current, frequency shift and third harmonic $a_3$ for a CuOx tip over a distance range of 50\,pm from (a) to (f) with decrements of 5\,pm. At $z=50$\,pm, the frequency shift image is close to the total charge density of graphene, i.e. a natural image of the surface. For closer distances, distortions start to appear before a pronounced asyammetry between A and B sites emerges at distances at and below $z=10$\,pm. Among the higher harmonics, only $a_3$ rises above the experimental noise, yet without a clear signature \cite{note_hh_measurement}.}\label{fig17_CHset_CuOx_I_df_hh}
\end{figure}

Figure \ref{fig17_CHset_CuOx_I_df_hh} shows again a constant height set where the tunneling current and frequency shift data look similar to the data set taken with a CO terminated tip of Fig. \ref{fig08_CHset_CO_I_df_hh}, yet they do not show strong third harmonics. However, the emergence of a strong A-B site difference becomes apparent for heights below 20\,pm from Figure \ref{fig17_CHset_CuOx_I_df_hh} (d).

\begin{figure}
\begin{center}
\includegraphics[clip=true, width=1\textwidth]{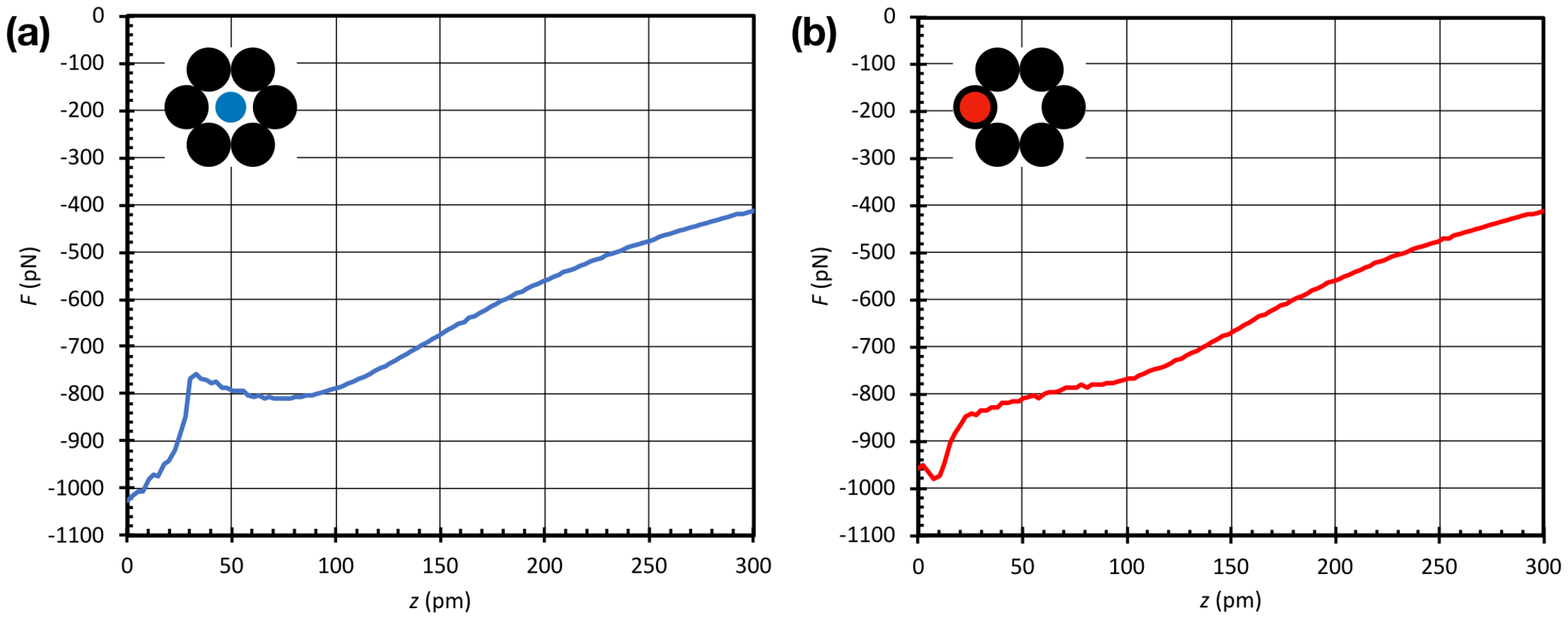}
\end{center}
\caption{Force versus distance spectra for a CuOx tip over a hollow site (a) and a carbon atom (b). (a) The sharp fall of the spectrum taken at the hollow site at $z=30$\,pm points to a possible onset of a graphite to diamond transition. (b) When taking the spectrum over a carbon atom, we expect that both sample- and tip deformations lead to a lateral displacement, finally ending with the oxygen front atom in the hollow site, causing a delayed force induced transition.}\label{fig18_F_CuOx_spectra}
\end{figure}

Also, the force versus distance curve over the hollow site displayed in Figure \ref{fig18_F_CuOx_spectra} (a) shows a pronounced step at a $z$ value of 30\,pm that points to a collapse of the lattice and a possible start of rehybridization. When performing the spectrum over an atom site in (b), the transition is softer and at closer $z$ values as expected.

\section{Images of graphene under weak repulsive forces for CO-terminated and CuOx tips}
\begin{figure}
\begin{center}
\includegraphics[clip=true, width=1\textwidth]{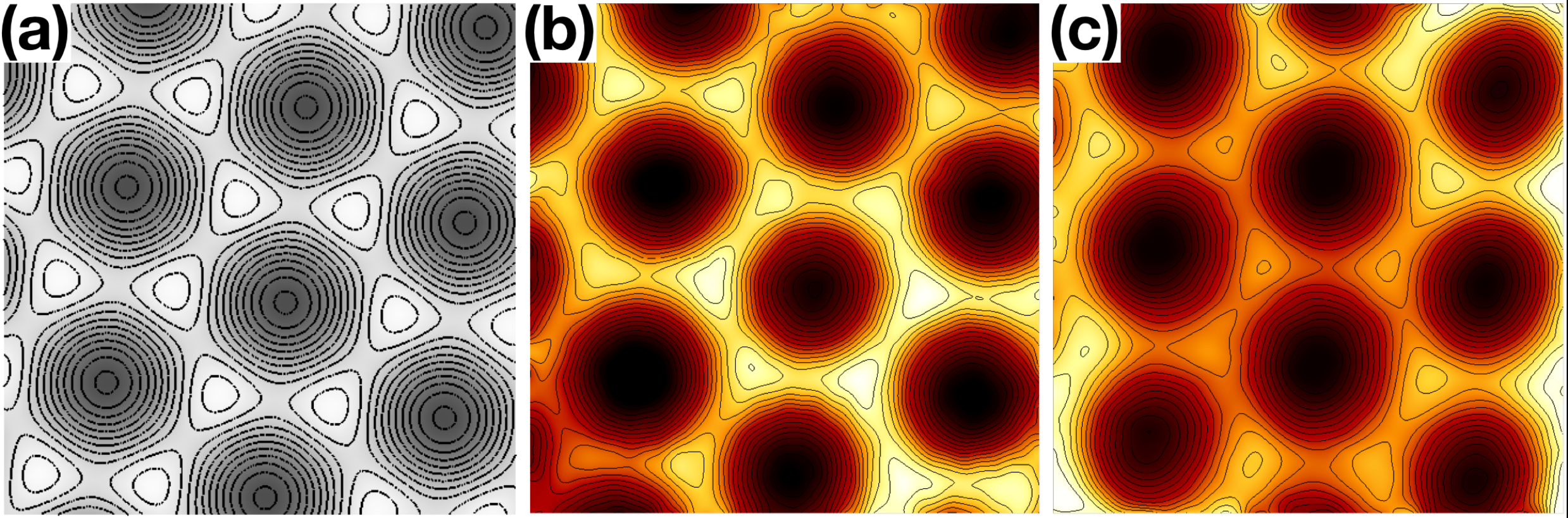}
\end{center}
\caption{Calculation of total charge density for graphene (a) and high resolution constant height images of graphene using a CO terminated tip (b) and a CuOx tip (c). Image size 0.6\,nm by 0.6\,nm. Parameters: (a) the total charge density is calculated at a height of 200\,pm above the plane that connects the nuclei of the carbon atoms. The charge density ranges from 2 (dark) to 4 (bright) electrons per nm$^3$, the contour lines are spaced by 0.2 electrons per nm$^3$. (b) Sensor type S1.0 (see \cite{Giessibl2019RSI} for specifications) with $k = 1\,800$\,N/m, $f_0 = 26\,660.3$\,Hz, $A = 50$\,pm and $Q=42\,844$. The contour lines are spaced by 0.1\,Hz, the frequency shift ranges from -8.2\,Hz (dark) to -4.9\,Hz (bright). 
(c) Sensor type M4 (see \cite{Giessibl2019RSI} for specifications) with $k = 1\,850$\,N/m, $f_0 =46\,597.3$\,Hz, $A = 50$\,pm and $Q=482\,321$.  
The contour lines are spaced by 0.1\,Hz, the frequency shift ranges from -33.5\,Hz (dark) to -30.2\,Hz (bright). 
}\label{fig19_g_CO_CuOx_highres}
\end{figure}

In scanning transmission electron microscopy, progress in collimating the electron beam to a diameter close to 130 pm enabled the resolution of the atomic lattice and defects in graphene \cite{Emtsev2009}. Graphene on SiC has been studied before by FM-AFM in vacuum at low temperatures \cite{Boneschanscher2012}, room temperature \cite{Telychko2015} and in ambient conditions \cite{Wastl2013}. The data here observed in a slight repulsive regime reveals a very subtle detail. Figure \ref{fig19_g_CO_CuOx_highres} (a) shows a charge density calculation of graphene at a height of 200\,pm over the planes connecting the C nuclei. Slater-type orbitals were used for the calculation as in \cite{Hembacher2003}, showing weak but noticable local maxima over the atom sites. When operating AFM with CO or CuOx tips at small repulsive forces, the experimental images clearly show the local maximum of charge density over the carbon atoms as displayed in Fig. \ref{fig19_g_CO_CuOx_highres} (b) for CO terminated tips and in (c) for CuOx tips. The overall contrast is similar for CO and CuOx tips, but the absolute value of the frequency shift is more negative for CuOx tips due to the larger vdW attraction.

\section{Summary}

To summarize the unification of the experimental observation of higher harmonics and the DFT data, we can explain the higher harmonics as a consequence of local and temporary transtion of graphene layers into a diamond structure. However, we can only explain the fivefold increase of van-der-Waals attraction to values up to 10\,nN by a runaway increase of vdW attraction due to indented graphene snuggling up closely against a relatively blunt metal tip. We also note that while the data was perfectly reproducible with that one tip, other tips that were possibly sharper did not produce the higher harmonics. Early data from 2004 \cite{Hembacher2004}, a time where tip preparation of AFMs was much less sophisticated and tips were known to be very blunt indicated by large negative frequency shifts showed even much higher magnitudes of higher harmonics, pointing to the importance of large runaway vdW attraction for the induction of a graphite to diamond transition and the subsequent generation of higher harmonics.

We note that as we do not yet have a recipe on how to prepare tips that reproduce the observed data, we cannot consider our data to be a proof for our suggested interpretation. In principle, some very odd tip artifact might also cause strong $a_3$'s. However, we routinely record higher harmonics on our low temperature microscopes and we have not ever observed such clear and strong higher harmonic signals with other CO terminated tips. 
One of the reviewers suggested that a tip that slides out of the local energy minimum in the center of the graphene hexagon might result in a force curve similar to the one plotted in Figure \ref{fig04_df_a2_to_a4} (c). Also, one could speculate that a breaking of the bond between the CO molecule and the metallic front atom, followed by rebonding to a metal atom in the second atomic layer of the tip might result in a similar force curve. However, we doubt that the current and frequency shift data would continue in such a smooth way over distance reduction in Figures \ref{fig08_CHset_CO_I_df_hh}, \ref{fig12_hh_ring_dia} and \ref{fig15_hh_dissipation}.

In summary, we have found strong additonal evidence to references \cite{Gao2018} and \cite{Cellini2021} that the tip of an AFM can cause a pressure induced transition from an sp$^2$ type bonding symmetry to a diamond-like sp$^3$ type bonding. The picometer scale study allows us to describe this process on the atomic scale, opening up high-pressure material science to the atomic dimension.

Density functional theory (DFT) calculations show that inducing a simultaneous transformation of mesoscale sheets of graphite into diamond by pushing with one CO tip on every $1\times 1$ unit cell  requires a pressure of 41\,GPa. However, the average pressure reduces to 36\,GPa for a graphene mesh with a $2\times 2$ unit cell and to 20\,GPa for a graphene mesh with a $2\sqrt{3}\times 2\sqrt{3}$ unit cell. From this observation of decreased transition pressures for larger unit cells, it appears reasonable that a large enough unit cell would indeed require merely 10\,GPa to induce the transition. This data is consistent with the nucleation mechanism proposed by Khaliullin et al. \cite{Khaliullin2011} where the artificial diamond creation does not involve the simultaneous transition of large sheets of graphite into diamond simultaneously, but by local transitions that occur sequentially. In addition, we expect that the non-adiabatic parts in the cantilever oscillation mimic high temperature effects.

\section{Appendix} 

\subsection{Combined STM and AFM with reactive single atom metal tips}
\label{subsec:Combined STM and AFM with reactive single atom metal tips}

The tip is crucial for image contrast in STM and AFM. It appears to be sensible to use a sharp single atom metal tip. However, Ondracek et al. \cite{Ondracek2011} calculated that single metal tips are very reactive even with respect to graphene and a tungsten tip can form a chemical bond with graphene where the attractive force reaches 2\,nN \cite{Ondracek2011}. This may come as a surprise, as graphite is a widely used lubricant. The graphene layers in graphite slide with very low friction on each other, even leading to superlubricity when those graphene layers are rotated as reported by Dienwiebel et al. \cite{Dienwiebel2004}.

Careful experiments have shown that the highly reactive single atom metal tips allow to obtain atomically resolved images as depicted in Figure \ref{fig20_sam_tip}. At large distance, the attractive forces to the tip are small enough such that atomically resolved STM data is obtainable in Figure \ref{fig20_sam_tip} (a), while the AFM data in (b) does not show atomic contrast yet.
Figures \ref{fig20_sam_tip} (c,d,g,h) record an event, where the graphene surface layer apparently moves about 150\,pm closer to the tip as indicated by the 30-fold jump in current in Fig. \ref{fig20_sam_tip} (h). At these close distances, atomic resolution  is observed in both the STM channel in Fig. \ref{fig20_sam_tip} (e) and the AFM channel in Fig. \ref{fig20_sam_tip} (f).

Recently, it was verified experimentally that single metal adatoms on a surface are more than twice as reactive as trimer clusters \cite{Berwanger2020}. Specifically, a single Fe atom on Cu(111) can exert up to 450\,pN of attraction to a CO terminated tip, while a Fe trimer only exerts a maximal force of 200\,pN \cite{Berwanger2020}. We expect a related effect for single atom versus trimer tips and the reduced reactivity of trimers versus monomers on surfaces is consistent with metallic trimer tips allowing atomic imaging of graphene, while single atom tips pose a danger to pull the graphene layers towards the tip.
\begin{figure}
\begin{center}
\includegraphics[clip=true, width=0.8\textwidth]{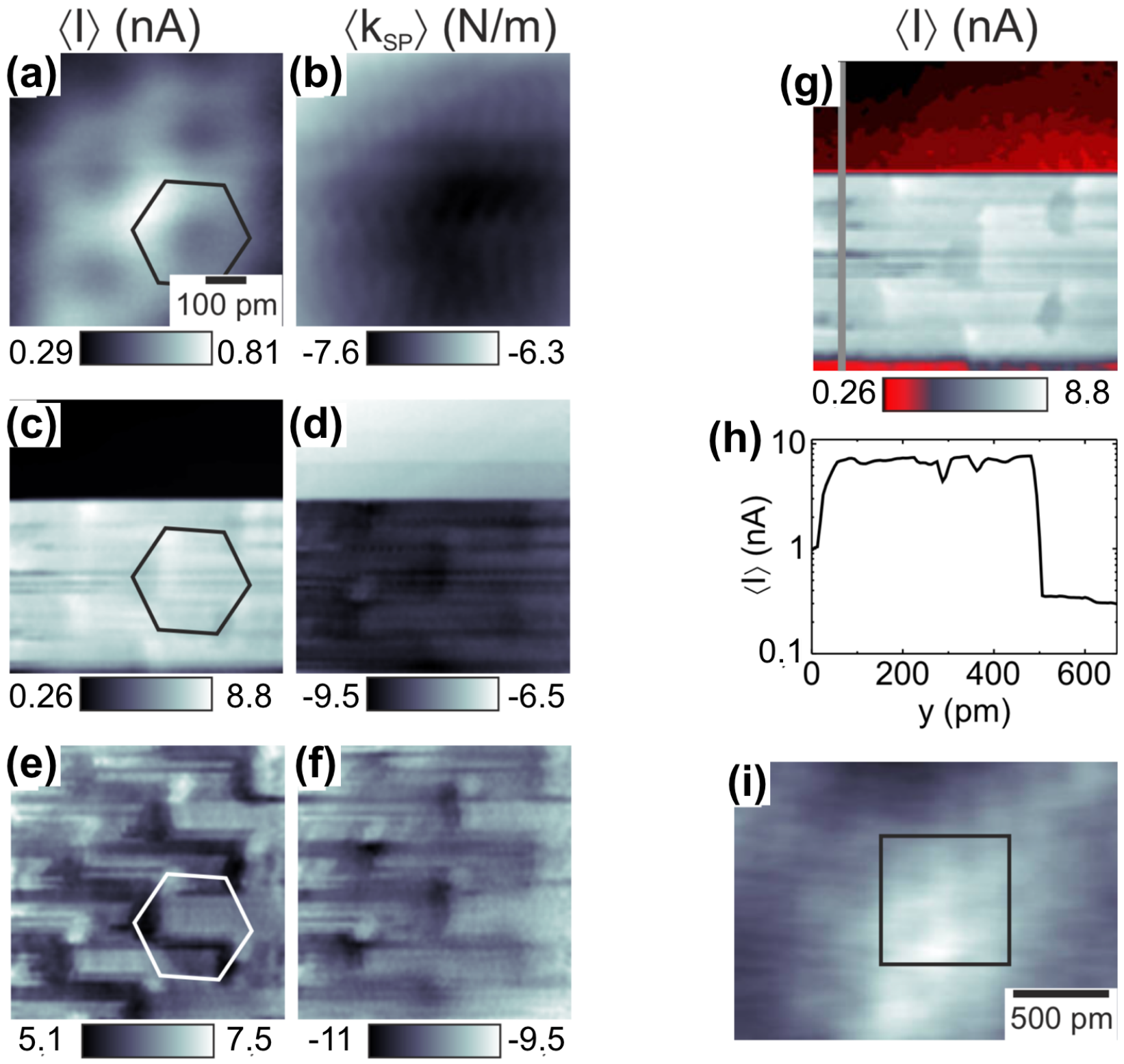}
\end{center}
\caption{STM and AFM data with a reactive single metal tip. (a) Atomically resolved STM data, while the AFM data in (b) does not provide atomic resolution. (c) Constant height STM data where the graphene flake jumps towards the tip by about 150\,pm, enabling atomic contrast in the AFM channel in (d) as well. (e) STM and (f) AFM data with atomic resolution with an unstable tip junction. (g) Constant height STM data and trace (h) showing a jump in current of more than 10, indicating a distance reduction of about 150\,pm. (i) Scan area of (a-g).}\label{fig20_sam_tip}
\end{figure}

\begin{figure}
\begin{center}
\includegraphics[clip=true, width=0.8\textwidth]{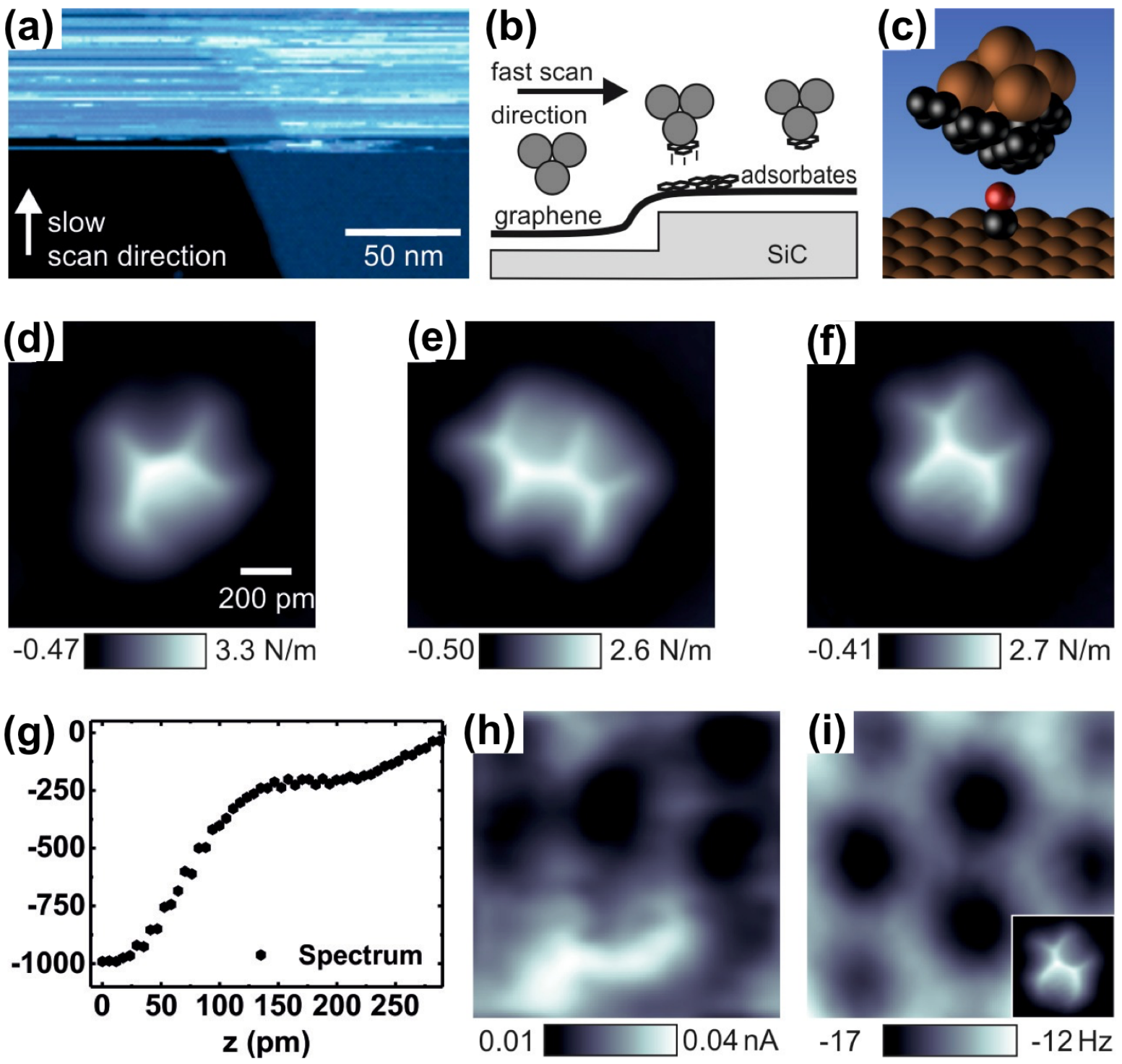}
\end{center}
\caption{(a) Topographic STM image of monolayer graphene with a reactive tip, where the graphene flake jumps to the tip (b) after about half the image was collected. (c) Sketch of STM tip covered by graphene. (d-f) Experimental tip images obtained by COFI (carbon monoxide front atom identification see ref. \cite{Welker2012}). The force versus distance spectrum (g) shows a shoulder that is caused by a lateral reversible drift of the tip during approach, allowing the tip to get even closer to reach a van-der-Waals attraction of about 1\,nN. STM (h) and AFM (i) images of graphene obtained with a graphene covered tip. The inset of (i) shows the COFI portrait of the tip used to obtain the image.}\label{fig21_g_gtip}
\end{figure}

Figure \ref{fig21_g_gtip} (a) shows a topographic STM image where a monolayer graphene on silicon carbide sample was scanned and the attractive forces between tip and sample were so large that the tip picked up a part of the graphene surface layer causing strong noise in the upper half of Figure \ref{fig21_g_gtip} (a). This pickup of graphene flakes by a reactive metal tip, illustrated in Figure \ref{fig21_g_gtip} (b), occured quite frequently and leads to a metal tip that has a graphene cover wrapped around it as shown in Figure \ref{fig21_g_gtip} (c).  

Inspection of these compromised tips by COFI \cite{Welker2012} is used to study and image the tip in Figure \ref{fig21_g_gtip} (d) after the event that occured in the midst of recording Figure \ref{fig21_g_gtip} (a). Similar events occured many times when starting with a monoatomic metal tip and the tips were always covered by graphene after these instabilties as shown in Figs. \ref{fig21_g_gtip} (e,f). While it was still possible to image graphene atomically with these graphene covered tips, the force versus distance spectra showed pronounced shoulders as evident in Figure \ref{fig21_g_gtip} (g). The appearance of the shoulder in Figure \ref{fig21_g_gtip} (g) is caused by a change in the dominating vertex of the graphene sheet that covers the tip induced by the reduction of distance, also accompanied by a large lateral shift of constant height data for small height variations (not shown here, see Fig. A.2 in \cite{Hofmann2014Diss} for details). Graphene terminated tips provided noisy images in the STM channel  (Fig. \ref{fig21_g_gtip} (h)) and also in the AFM channel (Fig. \ref{fig21_g_gtip} (i)), and cleaner data can be obtained with metallic trimer tips as shown in the following.

\subsection{Correction of CO bending}
\label{subsec:Correction of CO bending}

\begin{figure}
\begin{center}
\includegraphics[clip=true, width=0.8\textwidth]{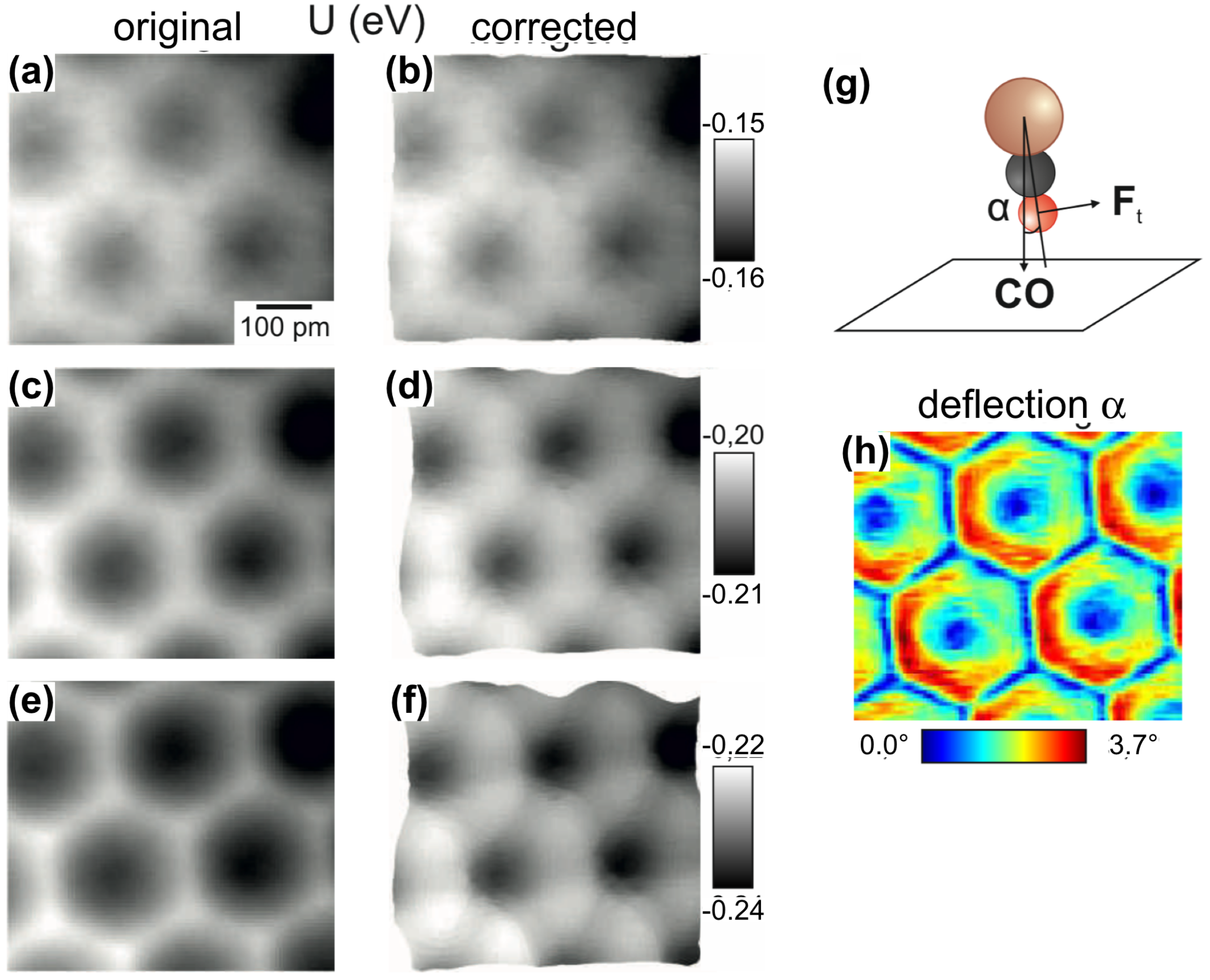}
\end{center}
\caption{Raw data (a,c,e) and corrected data (b,d,f) for short range potential energy plots deconvoluted from constant height frequency shift images. (g) The CO terminated tip is deflected laterally by an angle $\alpha$ that ranges from 0 to $3.7^\circ$ as shown in (h).}\label{fig22_bending_correction}
\end{figure}

A further demonstration of the effects of CO bending is provided by Figure \ref{fig22_bending_correction}, where we first use the normal force data from Figure \ref{fig22_bending_correction} (a-c) to compute the interaction energy between tip and sample and then take the lateral derivative to calculate the CO deflection and to correct the image \cite{Neu2014}. By utilizing the energy profile, calculating lateral forces from it and determining the lateral CO deflection \cite{Neu2014,Weymouth2014}, the CO deflection can be determined and the images can be corrected as shown in Fig. \ref{fig22_bending_correction}. 
The correction algorithms computes the lateral forces acting on the CO terminated probe tip, calculates its lateral bending and corrects the “true” lateral position of the image data, correcting the apparent bond sharpening and leading to warped edges of the corrected images. The correction algorithm shows that the apparent sharpening of the images for closer distances is an artifact caused by the lateral bending of the CO terminated tip. 

\subsection{Higher harmonics of a different order and illustration of higher harmonics generation}
\label{subsec:Higher harmonics of a different order and illustration of higher harmonics generation}

\begin{figure}
\begin{center}
\includegraphics[clip=true, width=0.8\textwidth]{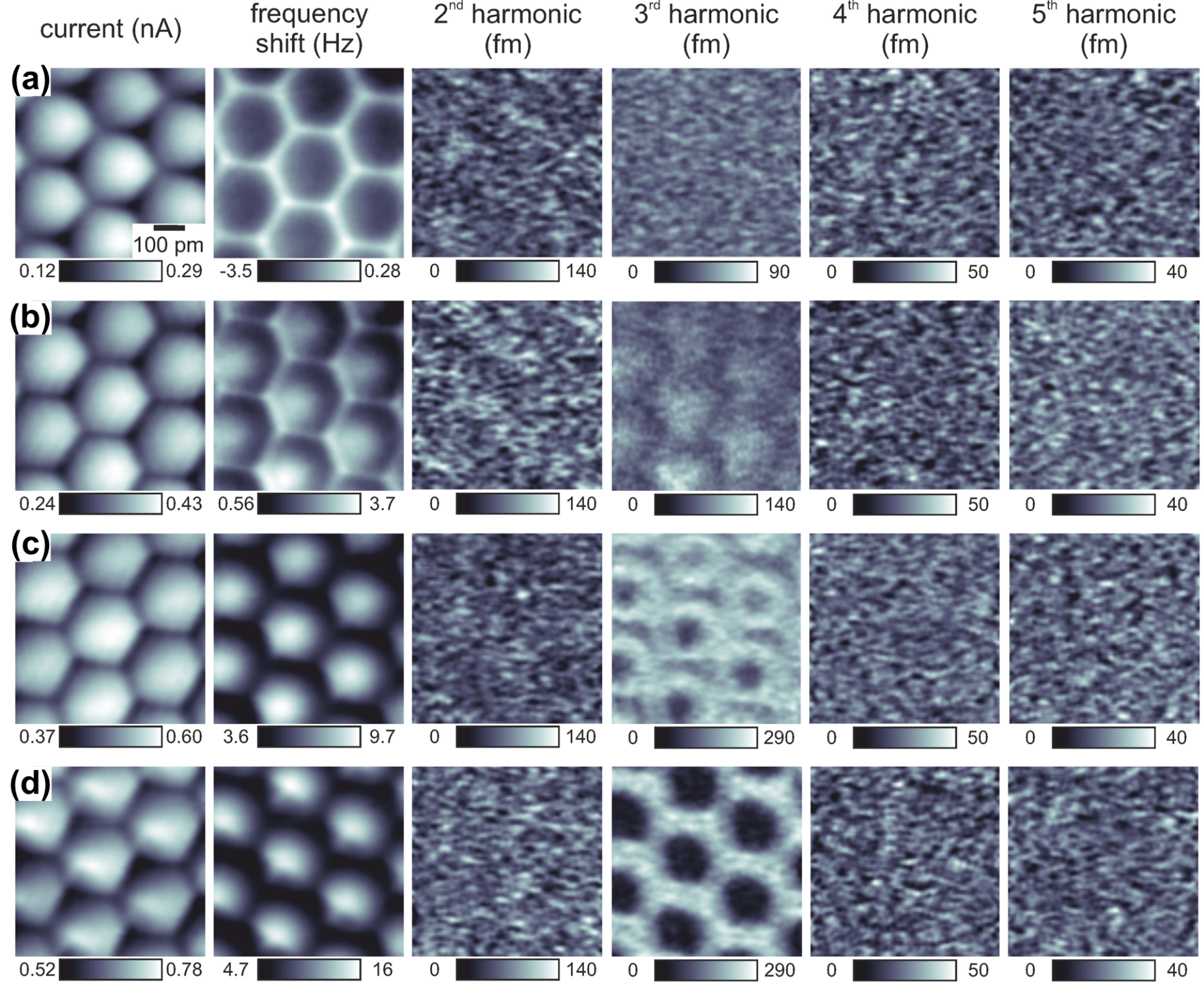}
\end{center}
\caption{Constant-height images of tunneling current, frequency shift and higher harmonics $a_2$, $a_3$, $a_4$, and $a_5$ for a CO terminated tip over a distance range of 48\,pm from (a) to (d) with decrements of 12\,pm.  Among the higher harmonics, $a_2$ may show a weak signal over the noise in (a-c), but $a_3$ rises very significantly above the experimental noise \cite{note_hh_measurement}. }\label{fig23_hh_2to5}
\end{figure}
One striking character of the higher harmonics is their very strong distance dependence. Another interesting property is that among the higher harmonics from $a_2$, $a_3$, $a_4$ and $a_5$, only $a_3$ rises significantly above the noise, as displayed in Fig. \ref{fig23_hh_2to5}. Although there is a weak signal in $a_2$ in (a), (b) and (c), only $a_3$ is large enough to allow for a detailed analysis.

\begin{figure}
\begin{center}
\includegraphics[clip=true, width=1\textwidth]{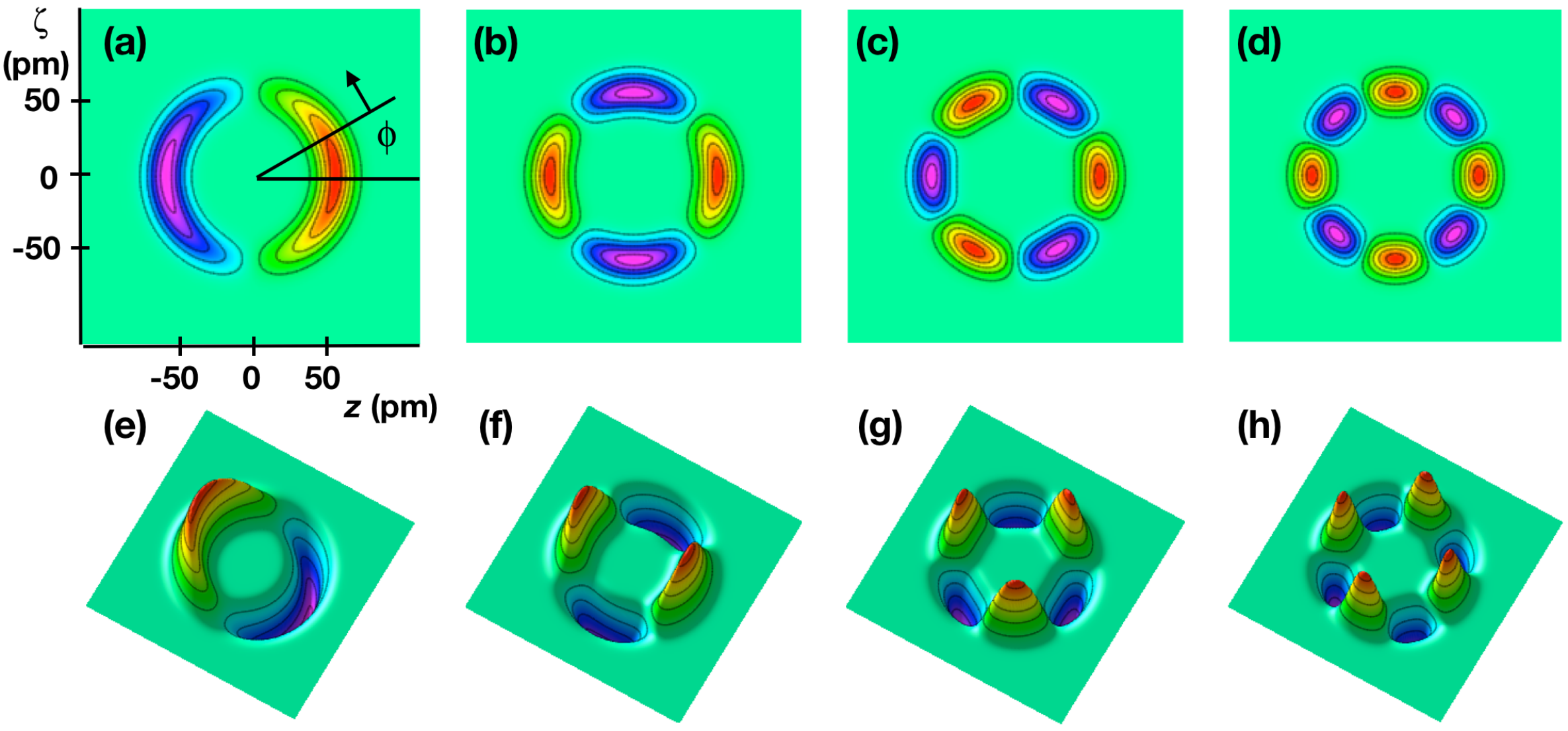}
\end{center}
\caption{Generation of higher harmonics $a_n$ of order $n$. The simple relation $a_n=F_{nf}/(k(1-n^2))$ where $F_{nf}$ is the Fourier component of the tip-sample force at frequency $n\cdot f$ yields a straighforward interpretation. -- Pseudo-3D representations of force fields $F(z)$ with $z=A\cos{\phi}$, $\phi=2\pi f t$ with oscillation frequency $f$ and time $t$ and $\zeta=A\sin{\phi}$ where the force is plotted in the vertical axis. 
(a) Force $F\propto z$ that generates a frequency shift, but no higher harmonics.
(b) Force $F\propto \cos(2\arccos(z/A))$ that generates a nonzero $a_2$, but neither a frequency shift nor higher harmonics of an order other than 2.
(c) Force $F\propto \cos(3\arccos(z/A))$ that generates a nonzero $a_3$, but neither a frequency shift nor higher harmonics of an order other than 3.
(d) Force $F\propto \cos(4\arccos(z/A))$ that generates a nonzero $a_4$, but neither a frequency shift nor higher harmonics of an order other than 4.
(e-h) 3D representations of the corresponding force fields $F(z)$.}\label{fig24_hh_gen}
\end{figure}

This particular representation of the higher harmonics allows for a descriptive graphic representation in Fig. \ref{fig24_hh_gen} - $a_{m}$ is the m-th Fourier coefficient of the force when expressed a function of $z+A\cos{\phi}$.

In particular, for $m=3$ we find:
\begin{equation}\label{a_m_det3}
a_{3}=-\frac{1}{8 k \cdot \pi } 
 \int_{0}^{2\pi}F_{ts}(z+A\cos{\phi})\cos (3\phi)d\phi.
\end{equation}
The magnitude of the higher harmonics is very small: for a force sensor with $k=1800$\,N/m, a third order amplitude $a_3=200$\,fm corresponds to a Fourier component of 2.88\,nN. This also implies that in practice, higher harmonics can only be detected at a very low detection bandwidth $B$. The noise in the higher harmonics is given by $\delta a_n = n_q(n f_0) \cdot \sqrt{B}$, where $n_q(f)$ is the frequency dependent deflection noise density that varies little with frequency (see section V in \cite{Giessibl2011PRB}) and is approximately 100\,fm/$\sqrt{\textrm{Hz}}$.

\subsection{Examples of imaging graphene with CO tips on other substrates}
\label{subsec:Examples of imaging graphene with CO tips on other substrates}

\begin{figure}
\begin{center}
\includegraphics[clip=true, width=0.8\textwidth]{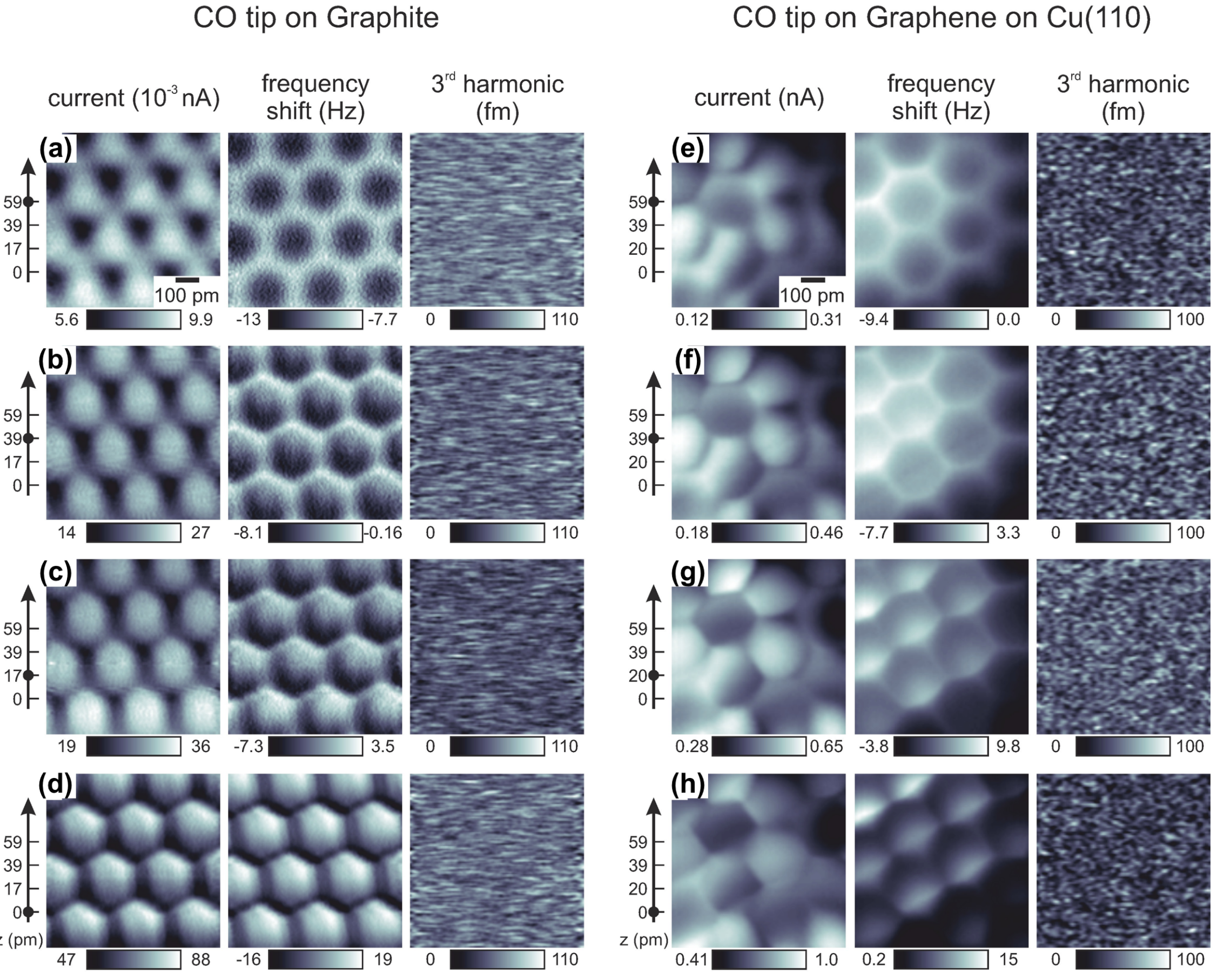}
\end{center}
\caption{(a-d) Tunneling current, frequency shift and third harmonic amplitude $a_3$ for highly oriented pyrolytic graphite (HOPG) for decreasing distances. (e-h)  Tunneling current, frequency shift and third harmonic amplitude $a_3$ for a small flake of graphene on Cu(110) for decreasing distances. The contrast of tunneling current (left column) and frequency shift (center column) shows a similar distance dependence as for graphene on SiC shown in Fig. \ref{fig08_CHset_CO_I_df_hh}. However, the higher harmonics $a_3$ remain below the noise level of about 100\,fm.}\label{fig25_g_HOPG_Cu110}
\end{figure}

In Figure \ref{fig25_g_HOPG_Cu110} (a-d), we look at HOPG (highly oriented pyrolytic graphite) with a CO tip and observe the same evolution in contrast for the tunneling current and frequency shift as observed for graphene on SiC above, but no higher harmonics. The same happens for graphene on Cu(110) as shown in Figure \ref{fig25_g_HOPG_Cu110} (e-h). 

\begin{figure}
\begin{center}
\includegraphics[clip=true, width=0.8\textwidth]{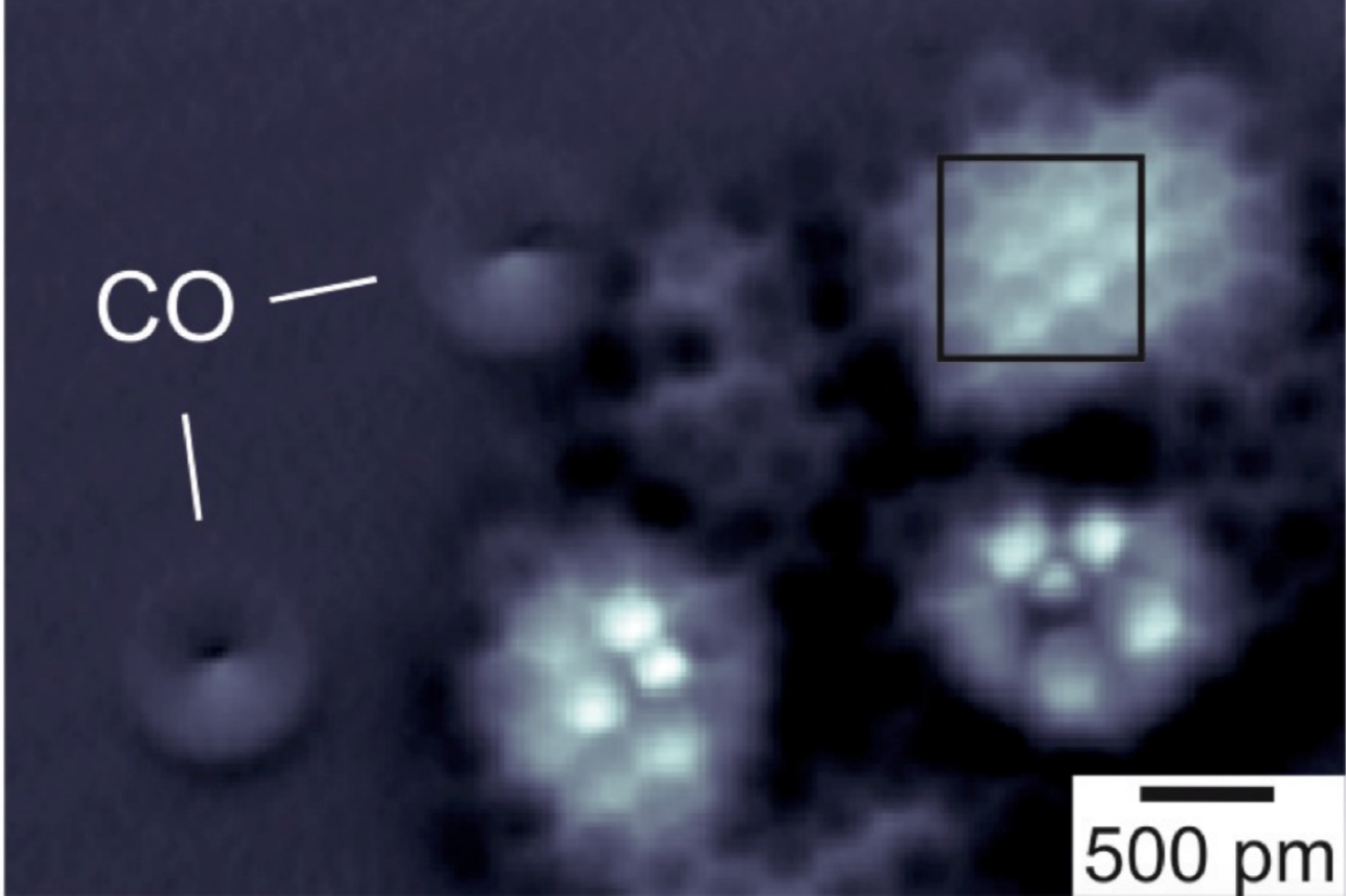}
\end{center}
\caption{Constant-height image of small graphene flakes that formed by thermal annealing of Cu(110), accompanied by two adsorbed CO molecules. The square frame shows the scan area imaged in Fig. \ref{fig25_g_HOPG_Cu110} (e-h).}\label{fig26_g_flake_Cu110}
\end{figure}

Figure \ref{fig26_g_flake_Cu110} (d) shows images of the graphene flakes with two random CO molecules that formed on Cu(110), where their small size explains why the STM data in Figure \ref{fig25_g_HOPG_Cu110} (e-h) is highly irregular, related to the irregularities that are observed in STM on multilayer graphene \cite{Stolyarova2007}.

\begin{figure}
\begin{center}
\includegraphics[clip=true, width=0.8\textwidth]{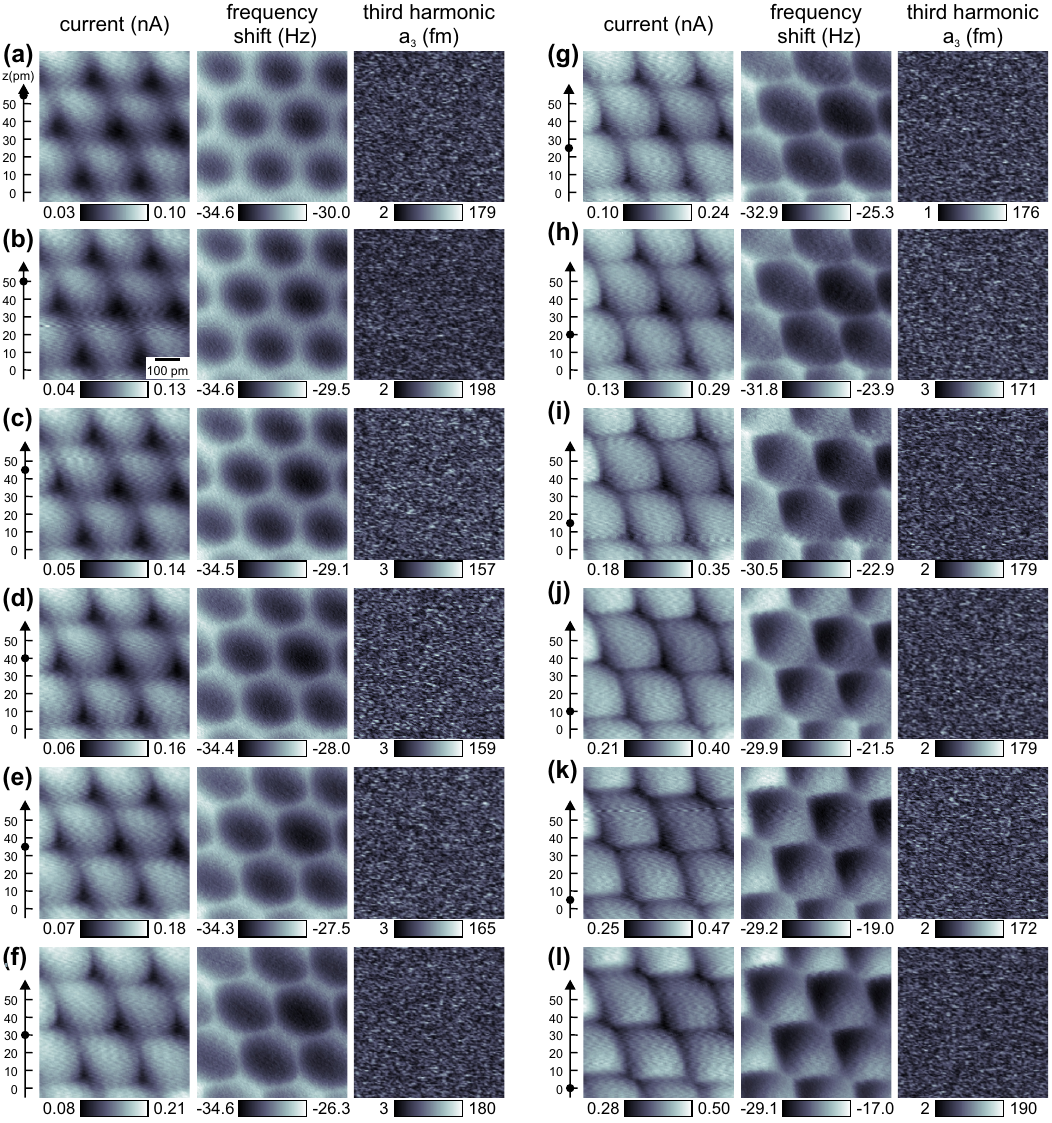}
\end{center}
\caption{Constant-height images of tunneling current, frequency shift and third harmonic $a_3$ for a CuOx tip over a distance range of 50\,pm from (a) to (f) with decrements of 5\,pm. At $z=50$\,pm, the frequency shift image is close to the total charge density of graphene, i.e. a natural image of the surface. For closer distances, distortions start to appear before a pronounced asyammetry between A and B sites emerges at distances at and below $z=10$\,pm.}\label{fig27_CuOx_ch_set}
\end{figure}

Figure \ref{fig27_CuOx_ch_set} is an extended constant height set of Fig. \ref{fig17_CHset_CuOx_I_df_hh} showing current, frequency shift and third harmonic over graphene with CuOx tips where the $z$ distance samples are spaced by merely 5\,pm.

\section{Remarks about the time sequence of these data} 

Indications for the proposed transition from sp$^2$ graphene to sp$^3$ diamond bonding induced by the pressure of an AFM tip may have been unknowingly observed decades ago. In 1987, Binnig observed the flipping rate of graphene in a low temperature STM operated in liquid helium \cite{Binnig1987}, showing strong A-B asymmetry of the flipping rate in Fig. 3 (b) of \cite{Binnig1987}. This could have been caused by a similar outward-motion of C atoms induced by ansp$^2$ to sp$^3$  transition as found here in Figure \label{fig20_sam_tip} (e,f,g). In 2004, Hembacher et al. \cite{Hembacher2004} used graphene to image a tungsten tip, where the front atom of a tungsten tip was imaged by either the A or B atoms of the graphite sample using strong creation of higher harmonics in the sensor oscillation. A local transition to diamond would explain the experimental observation of only one front atom image per unit cell of graphite, i.e. a dangling sp$^3$ bond that reaches from rehybridized B atoms into the vacuum. 

Stronger evidence for this transition was provided in the PhD thesis of Hofmann \cite{Hofmann2014Diss} (2014) where a well defined CO terminated tip \cite{Bartels1997,Gross2009} was used. An attempt to publish the data of \cite{Hofmann2014Diss} in a refereed journal failed. The reviewers praised the extraordinary quality of the experimental data, but felt that the interpretation leaves some open questions. In January 2016, one of the authors (FJG) presented the experimental data and the graphite-to-diamond transition hypothesis at the International Symposium of Surface- and Nanoscience (Furano, Japan). Matthias Scheffler from the Fritz-Haber-Institute Berlin, who had previously studied graphene on SiC with his team by density functional theory (DFT) \cite{Nemec2013}, also attended the meeting and proposed to investigate the rehybridization hypothesis by DFT. The calculations were performed by Xinguo Ren, and the results presented here confirmed that it is possible to induce a local graphite to diamond transition by pressing a CO terminated tip with a force on the order of 10\,nN onto the surface. However, the CO molecule in the DFT calculation had to be frozen in a vertical orientiation to prevent lateral bending when exerting this large repulsive force. Experimental evidence provided here shows that when approaching the CO tip to graphene, the CO automatically centers such that the O end of the tip finds a local minimum inmidst of the graphene hexagons when large repulisive forces act. 

The experimental observation of a strong third harmonic that marks overcoming a force threshold with subsequent lowering and increase of the force has only been observed with one CO terminated tip so far. Although this tip lasted for about 34 hours, generated about 900 images (64 constant height datasets with 7 data channels and two scan directions) and experienced  $\approx 9\times10^{9}$ approach-retract cycles over a $2A=100$\,pm distance range (oscillation at 26.6\,kHz for 34 hours), we did not succeed so far to prepare a CO terminated tip that showed the same resilience. 

As outlined further below, a CO terminated tip needs to fulfill two criteria to be able to induce a graphite-to-diamond transition and to generate a large third harmonic in the cantilever oscillation: 1. the CO terminated tip needs to sustain a repulsive force on the order of 10\,nN; 2. it needs to have a fairly flat metal base to generate a strong van-der-Waals attraction. A recent study has shown that CO terminated tips that are all formed by attaching CO to a single metallic front atom of the tip still show a significant variation in there force versus distance curves over metal clusters (Fig. 1 in \cite{Berwanger2020}).

\section{Acknowledgments} 
We thank Matthias Scheffler for very fruitful discussions based on his body of experience with performing DFT calculations for graphene on SiC and establishing contact with XR who performed the DFT calculations. 
We also thank Thomas Seyller for supplying samples, Lydia Nemec and Nikolai Moll for helpful discussions regarding DFT, and Florian Pielmeier for experimental contributions to the measurements of the graphene flake on Cu(110). Further, we thank Roland Bennewitz for providing very useful advice on the presentation of the experimental and theoretical data in the manuscript.
FJG thanks Roald Hoffmann for discussions that led to the inclusion of Figure \ref{fig19_g_CO_CuOx_highres} and Gerd Binnig for discussions regarding graphene reactivity. We thank the Deutsche Forschungsgemeinschaft for funding under GRK 1570.


\begin{thebibliography} {60}
\bibitem{note_hh_measurement} The HF2 Lock-in amplifier we used allows to record six different higher orders such as $a_2$ to $a_7$ in parallel, however the Nanonis BP4 scan controller only allowed to record four more channels (in addition to tunneling current, frequency shift, drive amplitude, …), therefore, Figure \ref{fig23_hh_2to5} only shows $a_2$, $a_3$, $a_4$ and $a_5$. We have checked higher orders than 5 manually, and they did also not exceed the noise level.
\end{thebibliography}
\end{document}